\documentclass{JHEP3}
\usepackage{amsmath}
\def\dslash#1{\setbox0=\hbox{$#1$}#1\hskip-\wd0\hbox to\wd0{\hss\sl/\/\hss}}
\makeatletter
\newcommand{\fmslash}[2][0mu]{%
  \mathchoice
    {\fmsl@sh\displaystyle{#1}{#2}}%
    {\fmsl@sh\textstyle{#1}{#2}}%
    {\fmsl@sh\scriptstyle{#1}{#2}}%
    {\fmsl@sh\scriptscriptstyle{#1}{#2}}}
\newcommand{\fmsl@sh}[3]{%
  \m@th\ooeqnarray{$\hfil#1\mkern#2/\hfil$\crcr$#1#3$}}
\makeatother

\title{A Supersymmetric Lagrangian for Fermionic Fields with Mass Dimension One}
\author{Kai E. Wunderle\thanks{Kai.Wunderle@usask.ca} and Rainer Dick\thanks{Rainer.Dick@usask.ca}\\ Department of Physics and Engineering Physics \\ University of Saskatchewan \\ 116 Science Place \\ Saskatoon, SK, S7N 5E2 \\ Canada} 
\abstract{We present the derivation of a supersymmetric model for fermionic fields with integer valued mass dimension based on a general superfield with one free spinor index.

First, we demonstrate that it is impossible to formulate such a model based on a general scalar superfield. This is due to problems constructing a Lagrangian containing a kinetic term for the fermionic mass dimension one field, as well as problems deriving a consistent second quantisation.

We then develop a formalism based on a general superfield with one free spinor index. We systematically derive all associated chiral and anti-chiral superfields up to third order in covariant derivatives. Using this formalism we are able to construct a supersymmetric on-shell Lagrangian that contains a kinetic term for the fermionic fields with mass dimension one. We then derive the corresponding on-shell supercurrent and succeed to formulate a consistent second quantisation for the component fields. Finally, we present our result for a supersymmetric Hamiltonian.

As the Lagrangian is by construction supersymmetric and the Hamiltonian was derived from the Lagrangian using the supersymmetry algebra the Hamiltonian must be positive definite.}
\preprint{}

\begin{document}
\section{Introduction}
All modern supersymmetric models are derived from a fundamental general scalar superfield. The application of the covariant superfield derivatives $D_\alpha$ and $\bar{D}_{\dot{\alpha}}$ allows then a systematic derivation of various chiral and anti-chiral superfields, see e.g. \cite{sohnius85, dine07}. If the discussion is restricted to the supersymmetric description of bosonic fields with integer-valued mass dimension and fermionic fields with half-integer-valued mass dimension this approach is sufficient and leads to the Minimal Supersymmetric Standard Model. This changes, however, if the previous assumption on the mass dimensions of fermionic and bosonic fields is dropped.

An investigation of this matter is of special interest due to the recent proposal of fermionic fields with mass dimension one -- eigenspinors of the charge conjugation operator (ELKO) -- by Ahluwalia-Khalilova and Grumiller \cite{ahluwaliakhalilova05a,ahluwaliakhalilova05b}. In their publications they use the field theory formalism to formulate a nonlocal theory of fermionic fields with mass dimension one. Ahluwalia et al. then modify this formalism by introducing a preferred direction along which the fermionic field with mass diemension one satisfies a local theory \cite{ahluwalia10}. Subsequently da Rocha and Hoff da Silva construct a Lagrangian for ELKO spinors motivated from supergravity using mass dimension transmuting operators \cite{darocha09}. Therfore, the question arises how to formulate a supersymmetric model from a fundamental superfield that is able to describe fermionic fields with mass dimension one.

The article is structured as follows. Section \ref{CLSUSY} discusses the construction of a supersymmetric Lagrangian. It is shown that the straightforward approach using a general scalar superfield with redefined mass dimensions fails while the construction of a model based on a general spinor superfield is successful. Then, the corresponding supercurrent is calculated in Section \ref{CJonshell}. In Section \ref{CHposition} the Hamiltonian is derived using the supersymmetry algebra. This approach ensures that the resulting Hamiltonian is positive definite. Finally, the results are summarised in Section \ref{Csummary}.

\section{A Supersymmetric Lagrangian}
\label{CLSUSY}
In this section a supersymmetric Lagrangian for fermionic fields with mass dimension one is derived. It is shown that a construction based on the general scalar superfield with redefined mass dimension of the component fields is impossible. This is due to problems generating a kinetic contribution for the fermionic fields with mass dimension one as well as constructing a consistent second quantisation. Afterwards the general spinor superfield is presented and all chiral and anti chiral superfields up to third order in covariant derivatives are derived systematically. This general spinor superfield is then used to construct a supersymmetric on-shell Lagrangian for fermionic fields with mass dimension one.
\subsection{Constructing a Model Based on the General Scalar Superfield}
\label{SLnot}
The most straightforward approach to formulate a supersymmetric model for fermionic fields with integer-valued mass dimension is to formulate a model in analogy to the commonly used formalism where fermionic fields have half-integer-valued mass dimension. This is done by starting from the general scalar superfield
\begin{align}
V
	&= C
	- i \theta \chi
	+ i \bar{\chi}' \bar{\theta}
	- \frac{i}{2} \theta^2 \left( M - i N \right)
	+ \frac{i}{2} \left( M + i N \right)
	- \theta \sigma^\mu \bar{\theta} A_\mu +\notag \\
	&\quad + i \bar{\theta}^2 \theta \left( \lambda - \frac{i}{2} \dslash{\partial} \bar{\chi}' \right)
	- i \theta^2 \bar{\theta} \left( \bar{\lambda}' - \frac{i}{2} \bar{\dslash{\partial}} \chi \right)
	- \frac{1}{2} \theta^2 \bar{\theta}^2 \left( D + \frac{1}{2} \Box C \right) ,
\end{align}
and redefining the mass dimensions of the component fields appropriately, e.g. $\mathrm{dim}{\left( C \right)} = 1/2$, $\mathrm{dim}{\left( \chi \right)} = 1$, etc. . The chiral superfields $X$ and $W_\alpha$ are then defined as
\begin{align}
X
	&= \frac{i}{2} \bar{D}^2 X \, ,\\
W_\alpha
	&= \frac{i}{4} \bar{D}^2 D_\alpha V \, .
\end{align}
where the covariant derivatives are given by
\begin{align}
D_\alpha
	&= \partial_\alpha - i \dslash{\partial}_{\alpha \dot{\beta}} \bar{\theta}^{\dot{\beta}} \, , \label{covariantD}\\
\bar{D}_{\dot{\alpha}}
	&= - \bar{\partial}_{\dot{\alpha}} + i \theta^\beta \dslash{\partial}_{\beta \dot{\alpha}} \, . \label{covariantDbar}
\end{align}
This choice of conventions differs by a factor of $- i$ from the conventions used in \cite{wess82}.

However, there are two fundamental problems that prevent a feasible theory using this approach. The first problem is that all possible contributions to the Lagrangian fail to produce a nonvanishing kinetic term for the fermionic fields. The second problem is encountered during second quantisation of the Lagrangian. It can be shown that already the simplest possible Lagrangian leads to negative energy solutions. In the following subsections these two problems will be discussed in detail.

\subsubsection{A Non-kinetic Supersymmetric Lagrangian}
\label{SSnonkinSUSYL}
\begin{TABLE}[t]{
\begin{tabular}{r|c|l}
Contribution & Mass Dimension & Possible Contributions \\
\hline
$V V$ & $\mathrm{dim}{\left( V V \right)} = 1$ & $\left( m V V \right)_D$ \\
\hline
$X V$ & $\mathrm{dim}{\left( X V\right)} = 2$ & $\left( X V \right)_D$ \\
$D V D V$ & $\mathrm{dim}{\left( D V D V\right)} = 2$ & $\left( D V D V \right)_D$  \\
$V X$ & $\mathrm{dim}{\left( V X \right)} = 2$ & $\left( V X \right)_D$ \\
\hline
$D W V$ & $\mathrm{dim}{\left( D W V \right)} = 3$ & mass dimension too big for D-component \\
$W D V$ & $\mathrm{dim}{\left( W D V \right)} = 3$ & mass dimension too big for D-component \\
$X X$ & $\mathrm{dim}{\left( X X \right)} = 3$ & $\left( X X \right)_F$ \\
$D V W$ & $\mathrm{dim}{\left( D V W \right)} = 3$ & mass dimension too big for D-component \\
$V D W$ & $\mathrm{dim}{\left( V D W \right)} = 3$ & mass dimension too big for D-component 
\end{tabular}
\caption{Contributions to the Lagrangian based on the general scalar superfield if $\chi$ is identified with the fermionic field of mass dimension one. In addition to the contributions built from products of unbarred superfields, the hermitian conjugates are permitted as well.}
\label{TChiDMnot}}
\end{TABLE}
The general scalar superfield has two possible candidates for a fermionic field with mass dimension one, $\chi$ and $\lambda$. For simplicity, the discussion is restricted to the case for $\chi$ as fermionic field with mass dimension one. Similar calculations can be repeated for $\lambda$. Due to the shift in mass dimension of the component fields the maximum number of covariant derivatives that needs to be considered is then increased by two and the discussion becomes more involved.

If $\chi$ is identified with the fermionic field with mass dimension one it can be shown that the mass dimensions of the general superfield $V$ and the chiral superfields $X$ and $W_\alpha$ are
\begin{align}
\mathrm{dim}{\left( V \right)}
	&= \frac{1}{2} \, , \quad
\mathrm{dim}{\left( X \right)}
	= \frac{3}{2} \, , \quad 
\mathrm{dim}{\left( W_\alpha \right)}
	= 2 \, . \quad
\end{align}
These results for the building blocks of the Lagrangian can be utilised to work out all possible contributions to the Lagrangian which have to satisfy three basic requirements. First, all contributions to the Lagrangian have to be Lorentz scalars and thus cannot contain any uncontracted indices. Second, all structure constants must have positive mass dimension for the theory to be renormalizable. Third, the contributions must have the appropriate mass dimension to contribute either via the $F$-component or the $D$-component.

All possible terms that satisfy the requirements are summarised in table \ref{TChiDMnot}. It groups the contributions into three groups depending on the mass dimension of the superfield product without structure constants. It is possible to conceive terms with higher mass dimension, however, those terms cannot contribute to the Lagrangian and are irrelevant for the following discussion. For simplicity the discussion is restricted to the unbarred fields while the hermitian conjugated components have to be considered for the Lagrangian as well.

The first group of terms with mass dimension one consists of one single term which is the product of two general superfields. As the general superfield is neither chiral nor antichiral the only possible contribution to the Lagrangian is a mass term via the D-component.

The second group containing all terms with mass dimension two then encompasses all terms that can be constructed using two general superfields and two covariant derivatives. This results in three possible contributions to the kinetic term via the D-component. There can be no contributions to the mass term via the F-component as neither $V$ nor $D V$ are chiral or anti-chiral.

Finally, the third group summarises all terms with mass dimension three which contain two general superfields as well as four covariant derivatives. Due to the mass dimension only contributions via the F-component are possible. The only term that satisfies the necessary symmetry requirements is $X X$ which contributes to the kinetic term.

Altogether, there is one contribution to the mass term as well as four contributions to the kinetic term. On the first glance this seems to ensure the existence of a valid model. However, explicit calculations reveal that neither of the four kinetic terms in question is able to produce a kinetic term for $\chi$ which was originally identified with the fermionic field with mass dimension one. A similar discussion can be repeated for the case where $\lambda$ is identified with the fermionic field with mass dimension one. Although the discussion for $\lambda$ produces an even larger number of potential contributions to the kinetic term neither of these terms produces a kinetic term for $\lambda$. Therefore, it can be concluded that it is impossible to construct a Lagrangian -- other than the trivial solution for a constant background spinor field -- based on the general scalar superfield that is able to describe fermionic fields with mass dimension one.

\subsubsection{Problems with Second Quantisation}
\label{SSProblemSQ}
The second major problem arises from the second quantisation of the component fields. A simple way to demonstrate this is to start with the simplest possible Lagrangian for a fermionic field with mass dimension one
\begin{align}
\mathcal{L}
	&= \partial_\mu \bar{\psi} \partial^\mu \psi
	- m^2 \bar{\psi} \psi \, .
\end{align}
The corresponding Hamiltonian is then found to be
\begin{align}
H
	&= \vec{\mathbf{\nabla}} \mathbf{\bar{\psi}} \vec{\mathbf{\nabla}} \mathbf{\psi}
	+ m^2 \bar{\psi} \psi \, .
\end{align}
Inserting the quantised Dirac field
\begin{align}
\psi
	&= \int \frac{\mathrm{d}^3 \mathbf{p}}{\left( 2 \pi \right)^3} \frac{1}{\sqrt{2 E_\mathbf{p}}} \sum\limits_s \left( a^s_\mathbf{p} u^s{\left( \mathbf{p} \right)} e^{-i p \cdot x} + {b^s_\mathbf{p}}^\dagger v^s{\left( \mathbf{p} \right)} e^{i p \cdot x} \right) \, , \\
\bar{\psi}
	&= \int \frac{\mathrm{d}^3 \mathbf{p}}{\left( 2 \pi \right)^3} \frac{1}{\sqrt{2 E_\mathbf{p}}} \sum\limits_s \left( b^s_\mathbf{p} \bar{v}^s{\left( \mathbf{p} \right)} e^{- i p \cdot x} + {a^s_\mathbf{p}}^\dagger \bar{u}^s{\left( \mathbf{p} \right)} e^{i p \cdot x} \right) \, ,
\end{align}
into the Hamiltonian and removing the zero-point energy leads to
\begin{align}
H
	&= \int \frac{\mathrm{d}^3 \mathbf{p}}{\left( 2 \pi \right)^3} m E_\mathbf{p} \sum\limits_s \left( {a^s_\mathbf{p}}^\dagger a^s_\mathbf{p} - {b^s_\mathbf{p}}^\dagger b^s_\mathbf{p} \right) .
\end{align}
The creation operator $b^\dagger$ can be used to lower the energy arbitrarily and obtain negative energy solutions.

\subsection{The General Superfield with one Spinor Index}
\label{SVgeneral}
In the previous section it was shown that a theory based on the general scalar superfield cannot be viable. This motivated an ansatz based on the general spinor superfield. So far only few references to the general spinor superfield exist in the literature. One exception being the article by Gates \cite{gates77} that contains an expansion of a spinor superfield in Grassmann variables. In addition, an expansion of the chiral spinor superfield was given by Siegel \cite{siegel79}. Their results are also included in the book by Gates, Grisaru, Ro\v{c}ek, and Siegel \cite{gates01}. As our notation differs from previous publications and is based on spinor superfields with different mass dimension the spinor superfield is introduced in detail and all chiral and anti-chiral superfields up to third order in covariant derivatives are derived.

In analogy to the general scalar superfield , the general spinor superfield can immediately be written down as expansion in $\theta$ and $\bar{\theta}$
\begin{align}
V_\alpha
	&= \kappa_\alpha
	+ \theta^\beta M_{\beta \alpha}
	+ \bar{\theta}^{\dot{\beta}} N_{\dot{\beta} \alpha}
	+ \theta^\beta \theta^\gamma \psi_{\alpha \beta \gamma}
	+ \bar{\theta}^{\dot{\beta}} \bar{\theta}^{\dot{\gamma}} \chi_{\alpha \dot{\beta} \dot{\gamma}}
	+ \theta^\beta \bar{\theta}^{\dot{\gamma}} \omega_{\alpha \beta \dot{\gamma}} + \notag \\
	&\quad + \theta^\beta \theta^\gamma \bar{\theta}^{\dot{\delta}} R_{\dot{\delta} \alpha \beta \gamma}
	+ \theta^\beta \bar{\theta}^{\dot{\gamma}} \bar{\theta}^{\dot{\delta}} S_{\alpha \beta \dot{\gamma} \dot{\delta}}
	+ \theta^\beta \theta^\gamma \bar{\theta}^{\dot{\delta}} \bar{\theta}^{\dot{\epsilon}} \lambda_{\alpha \beta \gamma \dot{\delta} \dot{\epsilon}} \, .
\end{align}
To bring this ansatz into a more convenient form the Grassmann variables need to be contracted over the respective indices. After absorbing some of the prefactors into the component fields, the general superfield is found to be
\begin{align}
V_\alpha
	&= \kappa_\alpha
	+ \theta^\beta M_{\beta \alpha}
	- \bar{\theta}^{\dot{\beta}} N_{\dot{\beta} \alpha}
	+ \theta^2 \psi_\alpha
	+ \bar{\theta}^2 \chi_\alpha
	+ \theta \sigma^\mu \bar{\theta} \left( \sigma_\mu \right)^{\beta \dot{\gamma}} \omega_{\alpha \beta \dot{\gamma}} - \notag \\
	&\quad - \theta^2 \bar{\theta}^{\dot{\delta}} R_{\dot{\delta} \alpha}
	+ \bar{\theta}^2 \theta^\beta S_{\beta \alpha}
	+ \theta^2 \bar{\theta}^2 \lambda_\alpha \, ,
\end{align}
where  $\kappa$, $\psi$, $\chi$, and $\lambda$ are Majorana spinors, $M$, $N$, $R$, and $S$ are complex second-rank spinors, and $\omega$ is a complex  third-rank spinor. The four complex second-rank spinors contain 32 bosonic degrees of freedom while the four Majorana spinors contain 16 fermionic degrees of freedom. As the number of bosonic and fermionic degrees of freedom must be the same for a supersymmetric theory, the 3-rd rank spinor must also have 16 fermionic degrees of freedom. It is then tempting to rewrite the third-rank spinor as a vector of majorana spinors
\begin{align}
\left( \sigma_\mu \right)^{\beta \dot{\gamma}} \omega_{\alpha \beta \dot{\gamma}}
	&= \left( \sigma_\mu \right)^{\beta \dot{\gamma}} \left( \sigma^\nu \right)_{\beta \dot{\gamma}} \omega_{\nu \alpha}
	= 2 \omega_{\mu \alpha} \, ,
\end{align}
which has 16 degrees of freedom as well. In the following discussion it will be referred to as a spinor-vector field. After appropriate rescaling of the component fields the most general spinor superfield is given by
\begin{align}
V_\alpha
	&= \kappa_\alpha
	+ \theta^\beta M_{\beta \alpha}
	- \bar{\theta}^{\dot{\beta}} N_{\dot{\beta}\alpha}
	+ \theta^2 \psi_\alpha
	+ \bar{\theta}^2 \chi_\alpha
	+ \theta \sigma^\mu \bar{\theta} \omega_{\mu \alpha} - \notag \\
	&\quad - \theta^2 \bar{\theta}^{\dot{\beta}} R_{\dot{\beta} \alpha}
	+ \bar{\theta}^2 \theta^\beta S_{\beta \alpha}
	+ \theta^2 \bar{\theta}^2 \lambda_\alpha \, .
\label{Valpha}
\end{align}

\subsubsection{The Chiral Superfields}
\label{SSChiralFields}
For the general scalar superfield the chiral and anti-chiral fields are derived by repeated operation of the covariant derivatives $D$ and $\bar{D}$. By definition the chiral and anti-chiral superfields satisfy the following relations
\begin{align}
\bar{D}_{\dot{\alpha}} X
	&= 0 \, , \\
D_\alpha Y
	&= 0 \, ,
\end{align}
where it is assumed that $X$ is a chiral superfield and $Y$ is an anti-chiral superfield which can have an arbitrary number of spinor indices. The chiral and anti-chiral superfields up to third order in covariant derivatives are then derived systematically by calculating $\bar{D}^2 V$ and $D^2 V$, as well as $\bar{D}^2 D V$ and $D^2 \bar{D} V$.

The chiral spinor field is found by repeated operation of the covariant derivative $\bar{D}$ onto the general superfield
\begin{align}
X_\alpha
	&= - \frac{1}{4} \bar{D}^2 V_\alpha \notag \\
	&= \chi_\alpha
	+ \theta^\beta \left( S_{\beta \alpha} + \frac{i}{2} \dslash{\partial}_\beta{}^{\dot{\delta}} N_{\dot{\delta} \alpha} \right)
	+ \theta^2 \left( \lambda_\alpha + \frac{i}{2} \partial^\mu \omega_{\mu \alpha} - \frac{1}{4} \Box \kappa_\alpha \right)
	- i \theta \dslash{\partial} \bar{\theta} \chi_\alpha + \notag \\
	&\quad + \frac{i}{2} \theta^2 \bar{\theta}^{\dot{\gamma}} \bar{\dslash{\partial}}_{\dot{\gamma}}{}^\beta \left( S_{\beta \alpha} + \frac{i}{2} \dslash{\partial}_\beta{}^{\dot{\delta}} N_{\dot{\delta} \alpha} \right)
	- \frac{1}{4} \theta^2 \bar{\theta}^2 \Box \chi_\alpha \, .
\label{Xchiral}
\end{align}
Comparison with the general expansion of a chiral field with one spinor index leads to the very elegant expression
\begin{align}
X_\alpha
	&= \exp{\left( - i \theta \dslash{\partial} \bar{\theta} \right)} \left( \chi_\alpha + \theta^\beta \left( S_{\beta \alpha} + \frac{i}{2} \dslash{\partial}_\beta{}^{\dot{\delta}} N_{\dot{\delta} \alpha} \right) + \theta^2 \left( \lambda_\alpha + \frac{i}{2} \partial^\mu \omega_{\mu \alpha} - \frac{1}{4} \Box \kappa_\alpha \right) \right) \, .
\end{align}
As this notation is not used in the further discussion an explicit notation in exponential form is not given for $Y$, $Z$, $Z_0$, and $Z'$ but can be derived in a similar way.

The calculations for the anti-chiral spinor field can be performed in perfect analogy where the operation of the covariant derivatives $D$ on the general superfield replaces the operation of $\bar{D}$
\begin{align}
Y_\alpha
	&= - \frac{1}{4} D^2 V_\alpha \notag \\
	&= \psi_\alpha
	- \bar{\theta}^{\dot{\beta}} \left( R_{\dot{\beta} \alpha} + \frac{i}{2} \bar{\dslash{\partial}}_{\dot{\beta}}{}^\gamma M_{\gamma \alpha} \right)
	+ \bar{\theta}^2 \left( \lambda_\alpha + \frac{i}{2} \partial^\mu \omega_{\mu \alpha} - \frac{1}{4}\Box \kappa_\alpha \right)
	- i \theta \dslash{\partial} \bar{\theta} \psi_\alpha - \notag \\
	&\quad - \frac{i}{2} \theta^\gamma \bar{\theta}^2 \dslash{\partial}_\gamma{}^{\dot{\beta}} \left( R_{\dot{\beta} \alpha} + \frac{i}{2} \bar{\dslash{\partial}}_{\dot{\beta}}{}^\delta M_{\delta \alpha} \right)
	- \frac{1}{4} \theta^2 \bar{\theta}^2 \Box \psi_\alpha \, .
\label{Yantichiral}
\end{align}

To third order in covariant derivatives there is again one chiral and one anti-chiral superfield which are now second-rank spinor fields.  The chiral second-rank spinor field is found to be
\begin{align}
Z_{\gamma \alpha}
	&= - \frac{1}{4} \bar{D}^2 D_\gamma V_\alpha \notag \\
	&= \left( S_{\gamma \alpha} - \frac{i}{2} \dslash{\partial}_\gamma{}^{\dot{\beta}} N_{\dot{\beta} \alpha} \right)
	+ \theta^\beta \left( 2 \epsilon_{\beta \gamma} \lambda_\alpha + \left( \sigma^{\nu \mu} \right)_{\beta \gamma} \partial_\nu \omega_{\mu \alpha} + \frac{1}{2} \epsilon_{\beta \gamma} \Box \kappa_\alpha \right) - \notag \\
	&\quad - i \theta^2 \left( \dslash{\partial}_\gamma{}^{\dot{\beta}} R_{\dot{\beta} \alpha} - \frac{i}{2} \Box M_{\gamma \alpha} \right)
	- i \theta \dslash{\partial} \bar{\theta} \left( S_{\gamma \alpha} - \frac{i}{2} \dslash{\partial}_\gamma{}^{\dot{\beta}} N_{\dot{\beta} \alpha} \right) + \notag \\
	&\quad + \frac{i}{2} \theta^2 \bar{\theta}^{\dot{\delta}} \dslash{\partial}^\beta{}_{\dot{\delta}} \left( 2 \epsilon_{\beta \gamma} \lambda_\alpha + \left( \sigma^{\nu \mu} \right)_{\beta \gamma} \partial_\nu \omega_{\mu \alpha} + \frac{1}{2} \epsilon_{\beta \gamma} \Box \kappa_\alpha \right) - \notag \\
	&\quad - \frac{1}{4} \theta^2 \bar{\theta}^2 \Box \left( S_{\gamma \alpha} - \frac{i}{2} \dslash{\partial}_\gamma{}^{\dot{\beta}} N_{\dot{\beta} \alpha} \right) \, .
\label{Zchiral}
\end{align}
A special case arises if the two undotted indices of the second-rank spinor field are contracted. It is then reduced to a scalar superfield
\begin{align}
Z_0
	&= \mathrm{Tr}{\left( S + \frac{i}{2} \dslash{\partial} N \right)}
	- \theta^\beta \left( 2 \lambda_\beta + \sigma^{\nu \mu} \partial_\nu \omega_\mu + \frac{1}{2} \Box \kappa_\beta \right) 
	+ i \theta^2 \mathrm{Tr}{\left( \dslash{\partial} R + \frac{i}{2} \Box M \right)} - \notag \\
	&\quad - i \theta \dslash{\partial} \bar{\theta} \mathrm{Tr}{\left( S + \frac{i}{2} \dslash{\partial} N \right)}
	- \frac{i}{2} \theta^2 \bar{\theta}^{\dot{\delta}} \dslash{\partial}^\beta{}_{\dot{\delta}} \left( 2 \lambda_\beta + \sigma^{\nu \mu} \partial_\nu \omega_\mu + \frac{1}{2} \Box \kappa_\beta \right) - \notag \\
	&\quad - \frac{1}{4} \theta^2 \bar{\theta}^2 \Box \mathrm{Tr}{\left( S + \frac{i}{2} \dslash{\partial} N \right)} \, .
\label{Z0chiral}
\end{align}
The calculations for the anti-chiral second-rank spinor field are nearly identical and it is found to be
\begin{align}
Z'
	&= - \frac{1}{4} D^2 \bar{D}_{\dot{\gamma}} V_\alpha \notag \\
	&= \left( R_{\dot{\gamma} \alpha} + \frac{i}{2} \dslash{\partial}^\beta{}_{\dot{\gamma}} M_{\beta \alpha} \right)
	- \bar{\theta}^{\dot{\beta}} \left( 2 \epsilon_{\dot{\beta} \dot{\gamma}} \lambda_\alpha - \left( \bar{\sigma}^{\nu \mu} \right)_{\dot{\beta} \dot{\gamma}} \partial_\nu \omega_{\mu \alpha} + \frac{1}{2} \epsilon_{\dot{\beta} \dot{\gamma}} \Box \kappa_\alpha \right) + \notag \\
	&\quad + \bar{\theta}^2 \left( i \bar{\dslash{\partial}}_{\dot{\gamma}}{}^\beta S_{\beta \alpha} - \frac{1}{2} \Box N_{\dot{\gamma} \alpha} \right)
	+ i \theta \dslash{\partial} \bar{\theta} \left( R_{\dot{\gamma} \alpha} + \frac{i}{2} \dslash{\partial}^\beta{}_{\dot{\gamma}} M_{\beta \alpha} \right) + \notag \\
	&\quad + \frac{i}{2} \theta^\delta \bar{\theta}^2 \dslash{\partial}_\delta{}^{\dot{\beta}} \left( 2 \epsilon_{ \dot{\beta} \dot{\gamma}}\lambda_\alpha - \left( \bar{\sigma}^{\nu \mu} \right)_{\dot{\beta} \dot{\gamma}} \partial_\nu \omega_{\mu \alpha} + \frac{1}{2} \epsilon_{\dot{\beta} \dot{\gamma}} \Box \kappa_\alpha \right) - \notag \\
	&\quad - \frac{1}{4} \theta^2 \bar{\theta}^2 \Box \left( R_{\dot{\gamma} \alpha} + \frac{i}{2} \dslash{\partial}^\beta{}_{\dot{\gamma}} M_{\beta \alpha} \right) \, .
\label{Zprimeantichiral}
\end{align}
Unlike for the chiral second-rank spinor field, no special case exists for the anti-chiral second-rank spinor field. This is due to the odd number of dotted and undotted indices which makes it impossible to contract the indices to achieve a scalar superfield. At most it can be written as a product of a Pauli-matrix and a vector field.

\subsubsection{Unitary Supertranslations}
\label{SSSUSYtranslation}
For the later discussion of the supercurrent and the derivation of the second quantisation of the component fields the superfield variation of the general spinor superfield must be derived. The calculation follows the discussion for the general scalar superfield in \cite{dick09} and was adapted accordingly to compensate for the additional spinor index.

The starting point for the derivation of the behaviour of a superfield under unitary supertranslations is the definition of a superspace eigenstate
\begin{align}
\left| x_0 , \theta_0 , \bar{\theta}_0 \right\rangle \, ,
\end{align}
which has the eigenvalues
\begin{align}
x^\mu \left| x_0 , \theta_0 , \bar{\theta}_0 \right\rangle
	&= x_0 \left| x_0 , \theta_0 , \bar{\theta}_0 \right\rangle \, , \\
\theta_\alpha \left| x_0 , \theta_0 , \bar{\theta}_0 \right\rangle
	&= \theta_{0 \alpha} \left| x_0 , \theta_0 , \bar{\theta}_0 \right\rangle \, , \\
\bar{\theta}_{\dot{\alpha}} \left| x_0 , \theta_0 , \bar{\theta}_0 \right\rangle
	&= \bar{\theta}_{0 \dot{\alpha}} \left| x_0 , \theta_0 , \bar{\theta}_0 \right\rangle \, .
\end{align}
Here $\theta_\alpha$, $\bar{\theta}_{\dot{\alpha}}$, and $x^\mu$ are operators acting on the superspace eigenstate while the eigenvalues are denoted by a subscript $0$ for the original eigenstate and with a prime for the translated state. Therefore, a state that is shifted under unitary supertranslations can be written as
\begin{align}
\left| x' , \theta' , \bar{\theta}' \right\rangle
	&= \exp{\left( i a y \cdot  P + i b \zeta Q + i c \bar{Q} \bar{\zeta} \right)} \left| x_0 , \theta_0 , \bar{\theta}_0 \right\rangle \, ,
\end{align}
where the prefactors $a$, $b$, and $c$ still need to be determined. An arbitrary operator $\mathcal{O}$ acting on the shifted state can be expressend as
\begin{align}
\mathcal{O} \left| x' , \theta' , \bar{\theta}' \right\rangle
	&= \exp{\left( i a y \cdot P + i b \zeta Q + i c \bar{Q} \bar{\zeta} \right)} \exp{\left( - i a y \cdot  P - i b \zeta Q - i c \bar{Q} \bar{\zeta} \right)} \times \notag \\
	& \quad \times \mathcal{O} \exp{\left( i a y \cdot P + i b \zeta Q + i c \bar{Q} \bar{\zeta} \right)} \left| x_0 , \theta_0 , \bar{\theta}_0 \right\rangle \, .
\end{align}
Using the Cambell-Baker-Hausdorff formula
\begin{align}
e^{- i G \lambda} \mathcal{O} e^{i G \lambda}
	&= \sum\limits_{j} \left( - i \lambda \right)^j \stackrel{j}{\left[ \right.} \!\!\! \left. G , \mathcal{O} \right]
\end{align}
this product of operators can be decomposed into an infinite sum of commutators
\begin{align}
\mathcal{O} \left| x' , \theta' , \bar{\theta}' \right\rangle
	&= \exp{\left( i a y \cdot P + i b \zeta Q + i c \bar{Q} \bar{\zeta} \right)} \times \notag \\
	&\quad \times \sum\limits_{j} \left( - i \lambda \right)^j \stackrel{j}{\left[ \right.} \!\!\! \left. a y \cdot P + b \zeta Q + c \bar{Q} \bar{\zeta} , \mathcal{O} \right] \left| x_0 , \theta_0 , \bar{\theta}_0 \right\rangle \, .
\end{align}
To evaluate the commutators it proves useful to utilise the following three commutators and anticommutators
\begin{align}
\left\{ \partial_\beta , \theta_\alpha \right\}
	&= \epsilon_{\beta \alpha} \, , \\
\left\{ \partial_{\dot{\beta}} , \bar{\theta}_{\dot{\alpha}} \right\}
	&= \epsilon_{\dot{\alpha} \dot{\beta}} \, , \\
\left[ P_\nu , x_\mu \right]
	&= i \eta_{\nu \mu} \, .
\end{align}

For the operator $\theta$ acting on the translated eigenstate it is found that
\begin{align}
\theta_\alpha \left| x' , \theta' , \bar{\theta}' \right\rangle
	&= \exp{\left( i a y \cdot P + i b \zeta Q + i c \bar{Q} \bar{\zeta} \right)} \times \notag \\
	&\quad \times \sum\limits_{j} \left( - i \lambda \right)^j \stackrel{j}{\left[ \right.} \!\!\! \left. a y \cdot P + b \zeta Q + c \bar{Q} \bar{\zeta} , \theta \right] \left| x , \theta , \bar{\theta} \right\rangle \, ,
\end{align}
where the $j$-th commutator has to be derived recursively. Conveniently, the first commutator is given by
\begin{align}
\left[ a y^\mu P_\mu + b \zeta^\beta Q_\beta + c \bar{Q}_{\dot{\beta}} \bar{\zeta}^{\dot{\beta}} , \theta_\alpha \right]
	&= i b \zeta_\alpha \, .
\end{align}
This implies that the second commutator vanishes. Therefore, all higher order contributions to the infinite sum must vanish identically as well and the eigenvalue of the shifted state is
\begin{align}
\theta'_\alpha \left| x' , \theta' , \bar{\theta}' \right\rangle
	&= \left( \theta_{0 \alpha} + b \zeta_\alpha \right) \left| x' , \theta' , \bar{\theta}' \right\rangle \, .
\end{align}

A similar calculation can be repeated for the operator $\bar{\theta}$. It is found that
\begin{align}
\bar{\theta}_{\dot{\alpha}} \left| x' , \theta' , \bar{\theta}' \right\rangle
	&= \exp{\left( i a y \cdot P + i b \zeta Q + i c \bar{Q} \bar{\zeta} \right)} \times \notag \\
	&\quad \times \sum\limits_{j} \left( - i \lambda \right)^j \stackrel{j}{\left[ \right.} \!\!\! \left. a y \cdot P + b \zeta Q + c \bar{Q} \bar{\zeta} , \bar{\theta}_{\dot{\alpha}} \right] \left| x_0 , \theta_0 , \bar{\theta}_0 \right\rangle \, .
\end{align}
Again, the $j$-th commutator must be calculated recursively starting with the first order commutator
\begin{align}
\left[ a y \cdot P + b \zeta^\beta Q_\beta + c \bar{Q}_{\dot{\beta}} \bar{\zeta}^{\dot{\beta}} , \bar{\theta}_{\dot{\alpha}} \right]
	&= i c \bar{\zeta}_{\dot{\alpha}} \, .
\end{align}
Like in the previous case this result implies that the second commutator vanishes identically and the eigenvalue of the shifted state is 
\begin{align}
\bar{\theta}'_{\dot{\alpha}} \left| x' , \theta' , \bar{\theta}' \right\rangle
	&= \left( \bar{\theta}_{0 \dot{\alpha}} + c \bar{\zeta}_{\dot{\alpha}} \right) \left| x' , \theta' , \bar{\theta}' \right\rangle \, .
\end{align}

Finally, the behaviour of the eigenvalue of the operator $x^\mu$ is analysed
\begin{align}
x^\mu \left| x' , \theta' , \bar{\theta}' \right\rangle
	&= \exp{\left( i a y \cdot P + i b \zeta Q + i c \bar{Q} \bar{\zeta} \right)} \times \notag \\
	&\quad \times \sum\limits_{j} \left( - i \lambda \right)^j \stackrel{j}{\left[ \right.} \!\!\! \left. a y \cdot P + i b \zeta Q + i c \bar{Q} \bar{\zeta} , x^\mu \right] \left| x_0 , \theta_0 , \bar{\theta}_0 \right\rangle \, .
\end{align}
The first commutator is found to be
\begin{align}
\left[ a y^\nu P_\nu + b \zeta^\alpha Q_\alpha + c \bar{Q}_{\dot{\alpha}} \bar{\zeta}^{\dot{\alpha}} , x^\mu \right]
	&= i a y^\mu
	- b \zeta \sigma^\mu \bar{\theta}
	+ c \theta \sigma^\mu \bar{\zeta} \, .
\end{align}
On the first glance it seems as if the series expansion doesn't terminate after the first commutator. However, the explicit calculation of the second commutator reveals that it vanishes identically
\begin{align}
\stackrel{2}{\left[ \right.} \!\!\! \left. a y^\nu P_\nu + b \zeta^\alpha Q_\alpha + c \bar{Q}_{\dot{\alpha}} \bar{\zeta}^{\dot{\alpha}} , x^\mu \right]
	&= 0 \, .
\end{align}
This terminates the infinite series and the eigenvalue for the operator $x^\mu$ acting on the translated state is given by
\begin{align}
x^\mu \left| x' , \theta' , \bar{\theta}' \right\rangle
	&= \left( x_0^\mu + a y^\mu + i \left( b \zeta \sigma^\mu \bar{\theta}_0 - c \theta_0 \sigma^\mu \bar{\zeta} \right) \right) \left| x' , \theta' , \bar{\theta}' \right\rangle \, .
\end{align}

Combining the results for the operators $\theta$, $\bar{\theta}$, and $x^\mu$ yields a translated superspace eigenstate of
\begin{align}
\left| x' , \theta' , \bar{\theta}' \right\rangle
	&= \left| x_0 + a y_0 + i \left( b \zeta \sigma \bar{\theta}_0 - c \theta_0 \sigma \bar{\zeta} \right) , \theta_0 + b \zeta , \bar{\theta}_0 + c \bar{\zeta} \right\rangle \, ,
\end{align}
where the prefactors $a$, $b$, and $c$ are still arbitrary. As a convention it is assumed that the discussion is restricted to pure superspace translations for which the spatial translation vanishes and thus $a y_0 =0$. Furthermore, the translations of the superspace coordinates $\theta$ and $\bar{\theta}$ are chosen to be positive which results in $b = c = 1$. This results in a relation between the original and shifted state of the following form
\begin{align}
\left| x' , \theta' , \bar{\theta}' \right\rangle
	&= \left| x + i \left( \zeta \sigma \bar{\theta} - \theta \sigma \bar{\zeta} \right) , \theta + \zeta , \bar{\theta} + \bar{\zeta} \right\rangle
	= \exp{\left( i \zeta Q + i \bar{Q} \bar{\zeta} \right)} \left| x , \theta , \bar{\theta} \right\rangle \, ,
\end{align}
where the subscript $0$ was dropped as it is no longer necessary to distinguish between operators and eigenvalues. The eigenstate at the shifted superspace coordinates is expressed in terms of the superspace coordinates of the original superspace eigenstate. It can be seen that a superspace translation, unlike a translation of normal fields, not only induces a spatial translation, but also results in a shift of the superspace coordinates $\theta$ and $\bar{\theta}$.

Now that the behaviour of a superspace eigenstate under unitary supertranslation is known, the calculation of the translated general spinor superfield is straightforward
\begin{align}
V'_\alpha{\left( x , \theta , \bar{\theta} \right)}
	&= \left\langle x , \theta , \bar{\theta} \right| \exp{\left( i \zeta Q + i \bar{Q} \bar{\zeta} \right)} \left| V_\alpha \right\rangle \notag \\
	&= \left\langle x - i \left( \zeta \sigma \bar{\theta} - \theta \sigma \bar{\zeta} \right) , \theta - \zeta , \bar{\theta} - \bar{\zeta} \right| \left. V_\alpha \right\rangle \notag \\
	&= V_\alpha{\left( x - i \left( \zeta \sigma \bar{\theta} - \theta \sigma \bar{\zeta} \right) , \theta - \zeta , \bar{\theta} - \bar{\zeta} \right)} \, .
\end{align}
As for the superspace eigenstate, a unitary supertranslation acting on a general superfield induces a spatial translation as well as a shift of superspace coordinates. In terms of the component fields the translated superfield can then be written as
\begin{align}
V'_\alpha{\left( x , \theta , \bar{\theta} \right)}
	&= \kappa_\alpha{\left( x - i \left( \zeta \sigma \bar{\theta} - \theta \sigma \bar{\zeta} \right) \right)}
	+ \left( \theta^\beta - \zeta^\beta \right) M_{\beta \alpha}{\left( x - i \left( \zeta \sigma \bar{\theta} - \theta \sigma \bar{\zeta} \right) \right)} - \notag \\
	&\quad - \left( \bar{\theta}^{\dot{\beta}} - \bar{\zeta}^{\dot{\beta}} \right) N_{\dot{\beta}\alpha}{\left( x - i \left( \zeta \sigma \bar{\theta} - \theta \sigma \bar{\zeta} \right) \right)}
	+ \left( \theta - \zeta \right)^2 \psi_\alpha{\left( x - i \left( \zeta \sigma \bar{\theta} - \theta \sigma \bar{\zeta} \right) \right)} + \notag \displaybreak[3] \\
	&\quad + \left( \bar{\theta} - \bar{\zeta} \right)^2 \chi_\alpha{\left( x - i \left( \zeta \sigma \bar{\theta} - \theta \sigma \bar{\zeta} \right) \right)} + \notag \displaybreak[3] \\
	&\quad + \left( \theta - \zeta \right) \sigma^\mu \left( \bar{\theta} - \bar{\zeta} \right) \omega_{\mu \alpha}{\left( x - i \left( \zeta \sigma \bar{\theta} - \theta \sigma \bar{\zeta} \right) \right)} - \notag \displaybreak[3] \\
	&\quad - \left( \theta - \zeta \right)^2 \left( \bar{\theta}^{\dot{\beta}} - \bar{\zeta}^{\dot{\beta}} \right) R_{\dot{\beta} \alpha}{\left( x - i \left( \zeta \sigma \bar{\theta} - \theta \sigma \bar{\zeta} \right) \right)} + \notag \displaybreak[3] \\
	&\quad + \left( \bar{\theta} - \bar{\zeta} \right)^2 \left( \theta^\beta - \zeta^\beta \right) S_{\beta \alpha}{\left( x - i \left( \zeta \sigma \bar{\theta} - \theta \sigma \bar{\zeta} \right) \right)} + \notag \\
	&\quad + \left( \theta - \zeta \right)^2 \left( \bar{\theta} - \bar{\zeta} \right)^2 \lambda_\alpha{\left( x - i \left( \zeta \sigma \bar{\theta} - \theta \sigma \bar{\zeta} \right) \right)} \, .
\label{SUSYvariationTaylor}
\end{align}
To express the translated component fields in terms of the component fields at the original superspace coordinates a Taylor expansion of the component fields can be used. In the present case an expansion up to first order in the transformation parameters $\zeta$ and $\bar{\zeta}$ of the form
\begin{align}
\kappa_\alpha{\left( x - i \left( \zeta \sigma \bar{\theta} - \theta \sigma \bar{\zeta} \right) \right)}
	&\approx \kappa_\alpha{\left( x \right)}
	- i \left( \zeta \sigma^\nu \bar{\theta} - \theta \sigma^\nu \bar{\zeta} \right) \partial_\nu \kappa_\alpha{\left( x \right)}
\end{align}
is sufficient. After appropriately rewriting equation (\ref{SUSYvariationTaylor}), neglecting all terms of second or higher order in the transformation parameters $\zeta$ and $\bar{\zeta}$, and collecting the terms with corresponding orders in the Grassmann variables $\theta$ and $\bar{\theta}$ the shifted superfield is given by
\begin{align}
V'_\alpha{\left( x , \theta , \bar{\theta} \right)}
	&= \kappa_\alpha{\left( x \right)}
	- \zeta^\beta M_{\beta \alpha}{\left( x \right)}
	+ \bar{\zeta}^{\dot{\beta}} N_{\dot{\beta}\alpha}{\left( x \right)} + \notag \\
	&\quad + \theta^\beta \left( M_{\beta \alpha}{\left( x \right)}
	+ i \left( \sigma^\mu \right)_{\beta \dot{\gamma}} \bar{\zeta}^{\dot{\gamma}} \partial_\mu \kappa_\alpha{\left( x \right)} 
	- 2 \zeta_\beta \psi_\alpha{\left( x \right)}
	- \left( \sigma^\mu \right)_{\beta \dot{\gamma}} \bar{\zeta}^{\dot{\gamma}} \omega_{\mu \alpha}{\left( x \right)} \right) - \notag \displaybreak[3] \\
	&\quad - \bar{\theta}^{\dot{\beta}} \left( N_{\dot{\beta}\alpha}{\left( x \right)}
	- i \left( \bar{\sigma}^\mu \right)_{\dot{\beta} \gamma} \zeta^\gamma \partial_\mu \kappa_\alpha{\left( x \right)}
	- 2 \bar{\zeta}_{\dot{\beta}} \chi_\alpha{\left( x \right)}
	- \left( \bar{\sigma}^\mu \right)_{\dot{\beta} \gamma} \zeta^\gamma \omega_{\mu \alpha}{\left( x \right)} \right) + \notag \displaybreak[3] \\
	&\quad + \theta^2 \left( \psi_\alpha{\left( x \right)}
	- \frac{i}{2} \bar{\zeta}^{\dot{\delta}} \left( \bar{\sigma}^\mu \right)_{\dot{\delta}}{}^\beta \partial_\mu M_{\beta \alpha}{\left( x \right)} 
	+ \bar{\zeta}^{\dot{\beta}} R_{\dot{\beta} \alpha}{\left( x \right)} \right) + \notag \displaybreak[3] \\
	&\quad + \bar{\theta}^2 \left( \chi_\alpha{\left( x \right)}
	- \frac{i}{2} \zeta^\delta \left( \sigma^\mu \right)_\delta{}^{\dot{\beta}} \partial_\mu N_{\dot{\beta}\alpha}{\left( x \right)} 
	- \zeta^\beta S_{\beta \alpha}{\left( x \right)} \right) + \notag \displaybreak[3] \\
	&\quad + \theta \sigma^\mu \bar{\theta} \left( \omega_{\mu \alpha}{\left( x \right)}
	+ \frac{i}{2} \zeta^\delta \left( \sigma^\nu \bar{\sigma}_\mu \right)_\delta{}^\beta \partial_\nu M_{\beta \alpha}{\left( x \right)}
	- \frac{i}{2} \bar{\zeta}^{\dot{\delta}} \left( \bar{\sigma}^\nu \sigma_\mu \right)_{\dot{\delta}}{}^{\dot{\beta}} \partial_\nu N_{\dot{\beta}\alpha}{\left( x \right)} - \right. \notag \displaybreak[3] \\
	&\qquad \left. - \zeta_\gamma \left( \sigma_\mu \right)^{\gamma \dot{\beta}} R_{\dot{\beta} \alpha}{\left( x \right)}
	+ \bar{\zeta}_{\dot{\gamma}} \left( \bar{\sigma}_\mu \right)^{\dot{\gamma} \beta} S_{\beta \alpha}{\left( x \right)} \right) - \notag \displaybreak[3] \\
	&\quad - \theta^2 \bar{\theta}^{\dot{\beta}} \! \left( \! R_{\dot{\beta} \alpha}{\left( x \right)} 
	- i \left( \bar{\sigma}^\mu \right)_{\dot{\beta} \gamma} \zeta^\gamma \partial_\mu \psi_\alpha{\left( x \right)}
	+ \frac{i}{2} \left( \bar{\sigma}^\mu \sigma^\nu \right)_{\dot{\beta} \dot{\epsilon}} \bar{\zeta}^{\dot{\epsilon}} \partial_\nu \omega_{\mu \alpha}{\left( x \right)}
	- 2 \bar{\zeta}_{\dot{\beta}} \lambda_\alpha{\left( x \right)} \! \right) \! + \notag \displaybreak[3] \\
	&\quad + \bar{\theta}^2 \theta^\beta \! \left( \! S_{\beta \alpha}{\left( x \right)}
	+ i \left( \sigma^\mu \right)_{\beta \dot{\gamma}} \bar{\zeta}^{\dot{\gamma}} \partial_\mu \chi_\alpha{\left( x \right)}
	+ \frac{i}{2} \left( \sigma^\mu \bar{\sigma}^\nu \right)_{\beta \delta} \zeta^\delta \partial_\nu \omega_{\mu \alpha}{\left( x \right)}
	- 2 \zeta_\beta \lambda_\alpha{\left( x \right)} \! \right) \! + \notag \\
	&\quad + \theta^2 \bar{\theta}^2 \left( \lambda_\alpha{\left( x \right)}
	- \frac{i}{2} \zeta^\gamma \left( \sigma^\mu \right)_{\gamma}{}^{\dot{\beta}} \partial_\mu R_{\dot{\beta} \alpha}{\left( x \right)}
	- \frac{i}{2} \bar{\zeta}^{\dot{\delta}} \left( \bar{\sigma}^\mu \right)_{\dot{\delta}}{}^\beta \partial_\mu S_{\beta \alpha}{\left( x \right)} \right) \, .
\label{SUSYvariation}
\end{align}
The variation of the general spinor superfield is then defined as the difference between the translated superfield and the superfield at the original superspace coordinates
\begin{align}
\delta V_\alpha{\left( x , \theta , \bar{ \theta} \right)}
	&= V'_\alpha{\left( x , \theta , \bar{ \theta} \right)} - V_\alpha{\left( x , \theta , \bar{ \theta} \right)} \, .
\end{align}
Therefore, the variation of the component fields can be extracted immediately from equation (\ref{SUSYvariation})
\begin{align}
\delta \kappa_\alpha
	&= - \zeta^\beta M_{\beta \alpha}{\left( x \right)}
	+ \bar{\zeta}^{\dot{\beta}} N_{\dot{\beta}\alpha}{\left( x \right)} \, , \label{deltakappa} \displaybreak[3] \\
\delta M_{\beta \alpha}
	&= - 2 \zeta_\beta \psi_\alpha{\left( x \right)}
	+ i \bar{\zeta}^{\dot{\gamma}} \left( \bar{\sigma}^\mu \right)_{\dot{\gamma} \beta} \partial_\mu \kappa_\alpha{\left( x \right)} 
	- \bar{\zeta}^{\dot{\gamma}} \left( \bar{\sigma}^\mu \right)_{\dot{\gamma} \beta} \omega_{\mu \alpha}{\left( x \right)} \, , \label{deltaM} \displaybreak[3] \\
\delta N_{\dot{\beta}\alpha}
	&= - 2 \bar{\zeta}_{\dot{\beta}} \chi_\alpha{\left( x \right)}
	- i \zeta^\gamma \left( \sigma^\mu \right)_{\gamma \dot{\beta}} \partial_\mu \kappa_\alpha{\left( x \right)}
	- \zeta^\gamma \left( \sigma^\mu \right)_{\gamma \dot{\beta}} \omega_{\mu \alpha}{\left( x \right)} \, , \label{deltaN} \displaybreak[3] \\
\delta \psi_\alpha
	&= \bar{\zeta}^{\dot{\beta}} R_{\dot{\beta} \alpha}{\left( x \right)} 
	- \frac{i}{2} \bar{\zeta}^{\dot{\beta}} \left( \bar{\sigma}^\mu \right)_{\dot{\beta}}{}^\gamma \partial_\mu M_{\gamma \alpha}{\left( x \right)} \, , \label{deltapsi} \displaybreak[3] \\
\delta \chi_\alpha
	&= - \zeta^\beta S_{\beta \alpha}{\left( x \right)}
	- \frac{i}{2} \zeta^\beta \left( \sigma^\mu \right)_\beta{}^{\dot{\gamma}} \partial_\mu N_{\dot{\gamma} \alpha}{\left( x \right)} \, , \label{deltachi} \displaybreak[3] \\
\delta \omega_{\mu \alpha}
	&= \zeta^\beta \left( \sigma_\mu \right)_\beta{}^{\dot{\gamma}} R_{\dot{\gamma} \alpha}{\left( x \right)}
	+ \frac{i}{2} \zeta^\beta \left( \sigma^\nu \bar{\sigma}_\mu \right)_\beta{}^\gamma \partial_\nu M_{\gamma \alpha}{\left( x \right)} - \notag \\
	&\quad - \bar{\zeta}^{\dot{\beta}} \left( \bar{\sigma}_\mu \right)_{\dot{\beta}}{}^\gamma S_{\gamma \alpha}{\left( x \right)} 
	- \frac{i}{2} \bar{\zeta}^{\dot{\beta}} \left( \bar{\sigma}^\nu \sigma_\mu \right)_{\dot{\beta}}{}^{\dot{\gamma}} \partial_\nu N_{\dot{\gamma}\alpha}{\left( x \right)} \label{deltaomega} \, , \displaybreak[3] \\
\delta R_{\dot{\beta} \alpha}
	&= - 2 \bar{\zeta}_{\dot{\beta}} \lambda_\alpha{\left( x \right)}
	- i \zeta^\gamma \left( \sigma^\mu \right)_{\gamma \dot{\beta}} \partial_\mu \psi_\alpha{\left( x \right)}
	- \frac{i}{2} \bar{\zeta}^{\dot{\gamma}} \left( \bar{\sigma}^\nu \sigma^\mu \right)_{\dot{\gamma} \dot{\beta}} \partial_\nu \omega_{\mu \alpha}{\left( x \right)} \, , \label{deltaR} \displaybreak[3] \\
\delta S_{\beta \alpha}
	&= - 2 \zeta_\beta \lambda_\alpha{\left( x \right)} 
	+ i \bar{\zeta}^{\dot{\gamma}} \left( \bar{\sigma}^\mu \right)_{\dot{\gamma} \beta} \partial_\mu \chi_\alpha{\left( x \right)}
	- \frac{i}{2} \zeta^\gamma \left( \sigma^\nu \bar{\sigma}^\mu \right)_{\gamma \beta} \partial_\nu \omega_{\mu \alpha}{\left( x \right)} \, , \label{deltaS} \displaybreak[3] \\
\delta \lambda_\alpha
	&= - \frac{i}{2} \zeta^\beta \left( \sigma^\mu \right)_{\beta}{}^{\dot{\gamma}} \partial_\mu R_{\dot{\gamma} \alpha}{\left( x \right)}
	- \frac{i}{2} \bar{\zeta}^{\dot{\beta}} \left( \bar{\sigma}^\mu \right)_{\dot{\beta}}{}^\gamma \partial_\mu S_{\gamma \alpha}{\left( x \right)} \, . \label{deltalambda}
\end{align}
These results then imply the variation of the on-shell component fields. After eliminating the auxiliary fields and using the definition of the component fields $\tilde{R}$ and $\tilde{S}$ from equations (\ref{tildeS}) and (\ref{tildeR}) in section \ref{SSLmassiveonshell} the variation of the component field of the on-shell Lagrangian are found to be
\begin{align}
\delta \psi_\alpha
	&= \bar{\zeta}^{\dot{\beta}} \tilde{R}_{\dot{\beta} \alpha} \, , \displaybreak[3] \label{deltapsionshell}\\
\delta \chi_\alpha
	&= - \zeta^\beta \tilde{S}_{\beta \alpha} \, , \displaybreak[3] \label{deltachionshell}\\
\delta \tilde{R}_{\dot{\beta} \alpha}
	&= m \bar{\zeta}_{\dot{\beta}} \chi_\alpha
	- 2 i \zeta^\gamma \dslash{\partial}_{\gamma \dot{\beta}} \psi_\alpha \, , \displaybreak[3] \label{deltaRonshell}\\
\delta \tilde{S}_{\beta \alpha}
	&= m \zeta_\beta \psi_\alpha
	+ 2 i \bar{\zeta}^{\dot{\gamma}} \bar{\dslash{\partial}}_{\dot{\gamma} \beta} \chi_\alpha \, . \label{deltaSonshell}
\end{align}

\subsection{Constructing a Model Based on the General Spinor Superfield}
\label{SSLchi}
\begin{TABLE}[t]{
\begin{tabular}{r|c|l}
Product & Mass Dimension & Contributions \\
\hline
$V V$ & $\mathrm{dim}{\left( V V \right)} = 0$ & $\left( m^2 V V \right)_D$ \\
\hline
$X V$ & $\mathrm{dim}{\left( X V \right)} = 1$ & $\left( m X V \right)_D$ , $\left( m Y V \right)_D$\\
$D V D V$ & $\mathrm{dim}{\left( D V D V \right)} = 1$ & $\left( m D V D V \right)_D$ , $\left( m \bar{D} V \bar{D} V \right)_D$ \\
$V X$ & $\mathrm{dim}{\left( V X \right)} = 1$ & $\left( m V X \right)_D$ , $\left( m V Y \right)_D$ \\
\hline
$D Z V$ & $\mathrm{dim}{\left( D Z V \right)} = 2$ & $\left( D Z V \right)_D$ , $\left( \bar{D} Z' V \right)_D$\\
$Z D V$ & $\mathrm{dim}{\left( Z D V \right)} = 2$ & $\left( Z D V \right)_D$ , $\left( Z' \bar{D} V \right)_D$\\
$X X$ & $\mathrm{dim}{\left( X X \right)} = 2$ & $\left( m X X \right)_F$ , $\left( m Y Y \right)_F$ , $\left( X Y \right)_D$ , $\left( Y X \right)_D$ \\
$D V Z$ & $\mathrm{dim}{\left( D V Z \right)} = 2$ & $\left( D V Z \right)_D$ , $\left( \bar{D} V Z' \right)_D$ \\
$V D Z$ & $\mathrm{dim}{\left( V D Z \right)} = 2$ & $\left( V D Z \right)_D$ , $\left( V \bar{D} Z' \right)_D$ \\
\hline
$D Z X$ & $\mathrm{dim}{\left( D Z X \right)} = 3$ & mass dimension too big for D-component \\
$Z Z$ & $\mathrm{dim}{\left( Z Z \right)} = 3$ & $\left( Z Z \right)_F$ , $\left( Z' Z' \right)_F$ \\
$X D Z$ & $\mathrm{dim}{\left( X D Z \right)} = 3$ & mass dimension too big for D-component
\end{tabular}
\caption{Possible contributions to the Lagrangian for $\chi$ as fermionic field with mass dimension one based on the general spinor superfield. The first two columns specify the product and mass dimensionality using the general superfield and chiral superfields only. The third column then summarises all possible contributions corresponding to the product outlined in the first column including the contributions that arise from the antichiral superfields.}
\label{TChiDM}}
\end{TABLE}
If $\chi$ is identified with the fermionic field with mass dimension one it can be shown that
\begin{align}
\mathrm{dim}{\left( V_\alpha \right)}
	&= 0 \, , \quad
\mathrm{dim}{\left( X_\alpha \right)}
	= \mathrm{dim}{\left( Y_\alpha \right)}
	= 1 \, , \quad
\mathrm{dim}{\left( Z_{\gamma \alpha} \right)}
	= \mathrm{dim}{\left( Z'_{\dot{\gamma} \alpha} \right)}
	= \frac{3}{2} \, .
\end{align}
It is interesting to note that for $\chi$ as fermionic field with mass dimension one the mass dimension of the general spinor superfield is $1/2$ lower than for the previous approach based on the the general scalar superfield. This indicates that the structure of this model is richer as there are more allowed contributions to the Lagrangian. For convenience the discussion is resticted to the unbarred superfields while the hermitian conjugates contribute to the Lagrangian as well.

The contributions to the Lagrangian have to satisfy the same basic requirements as outlined in Section \ref{SSnonkinSUSYL} -- no uncontracted spinor indices, positive mass dimension for structure constants, and appropriate mass dimension for contribution via $D$- or $F$-component. All conceivable terms that are in agreement with these conditions are then summarised in table \ref{TChiDM}. It contains more possible contributions to the Lagrangian which are now divided into four groups. The additional group is due to the lower mass dimensionality of the general spinor superfield which allows a spectrum for the mass dimension ranging from 0 and 3.

The first group which contains only one term, the product of two general spinor superfields without additional covariant derivatives, has mass dimension 0. For symmetry reasons the only possible contribution to the Lagrangian is a mass term via the D-component.

The second group containing all terms with mass dimension 1 has 6 possible terms. As $V$ and $D V$ are neither chiral nor anti-chiral all six terms are contributions to the mass term via the D-component.

In the third group all terms with mass dimension 2 are grouped together. It contains 12 terms of which 10 are contributions to the kinetic term via the D-component while 2 are contributions to the mass term via the F-component. It is worth mentioning that this is the only group that contains contributions to the kinetic term as well as contributions to the mass term. Even more intriguing is the fact that a superfield product of the form $X_1 X_2$ where $X_1$ and $X_2$ can be either chiral or antichiral is able to produce both kind of contributions.

Finally, the fourth group which contains all terms with mass dimension 3 has two entries. Due to the mass dimension only contributions via the F-component are possible which means that both terms can only contribute to the kinetic term.

It is interesting to note that some of the terms contained in table \ref{TChiDM}, namely $D V D V$ and $X V$ were previously considered by Gates and Siegel \cite{gates80,gates81}. However, in these articles the authors assume the commonly used mass dimensinos for fermionic and bosonic fields. This has two consequences. First, all kinetic terms in \cite{gates80,gates81} become mass terms in the present scenario due to the change of mass dimensionality. Second, all contributions summarised in groups three and four of table \ref{TChiDM}, and therefore the products of chiral superfields $X X$ and $Y Y$ do not exist without redefinition of mass dimensions to accommodate fermionic fields with mass dimension one and thus were not considered before.

\subsection{The On-shell Lagrangian}
\label{SSLmassiveonshell}
A supersymmetric Lagrangian can be constructed by combining contributions that were found in the dimensional analysis of the previous section. It was mentioned earlier that the first two groups of table \ref{TChiDM} with mass dimension 0 and 1 respectively contain only contributions to the mass term while the group with mass dimension 3 only produces contributions to the kinetic term. Therefore, the following discussion for the construction of a supersymmetric Lagrangian will be resticted to the third group which is the only one containing kinetic as well as mass terms. This limits the number of superfield products that need to be calculated to 12. Explicit calculations reveal that this number can be narrowed down even further. It can be shown that the terms $\left( D Z V \right)_D$, $\left( Z D V \right)_D$, $\left( X Y \right)_D$, $\left( D V Z \right)_D$, $\left( V D Z \right)_D$ are identical up to a prefactor. Therefore, only the D-component of the terms $X Y$ and $Y X$ will be considered for the kinetic term. The Lagrangian can then be written in a very compact form
\begin{align}
\mathcal{L}
	&= \left( X Y \right)_D + \left( Y X \right)_D + \frac{m}{2} \left( X X \right)_F + \frac{m}{2} \left( Y Y \right)_F + h.c. \, .
\label{Lcompact}
\end{align}
From the previous derivation of the chiral superfield $X$ in equation (\ref{Xchiral}) and the anti-chiral superfield $Y$ in equation (\ref{Yantichiral}) it can be seen that the component fields $N$, $M$, $S$, $R$, $\lambda$, and $\kappa$ are not independent. Therefore, it is convenient to introduce the new component fields
\begin{align}
\tilde{S}_{\beta \alpha}
	&= S_{\beta \alpha} + \frac{i}{2} \dslash{\partial}_\beta{}^{\dot{\gamma}} N_{\dot{\gamma} \alpha} \, , \label{tildeS} \\
\tilde{R}_{\dot{\beta} \alpha} 
	&= R_{\dot{\beta} \alpha} - \frac{i}{2} \bar{\dslash{\partial}}_{\dot{\beta}}{}^\tau M_{\tau \alpha} \, , \label{tildeR} \\
\tilde{\lambda}_\alpha
	&= \lambda_\alpha - \frac{1}{4} \Box \kappa_\alpha \, . \label{tildelambda}
\end{align}
Furthermore, it can be seen that the spinor-vector field $\omega^\mu_\alpha$ is always contracted with a four derivative which is simplified by defining
\begin{align}
\tilde{\omega}_\alpha 
	&= \partial^\mu \omega_{\mu \alpha} \, . \label{tildeomega}
\end{align}
The chiral and anti-chiral superfields can then be written as
\begin{align}
X_\alpha
	&= \chi_\alpha
	+ \theta^\beta \tilde{S}_{\beta \alpha}
	+ \theta^2 \left( \tilde{\lambda}_\alpha + \frac{i}{2} \tilde{\omega}_\alpha \right)
	- i \theta \dslash{\partial} \bar{\theta} \chi_\alpha
	+ \frac{i}{2} \theta^2 \bar{\theta}^{\dot{\gamma}} \bar{\dslash{\partial}}_{\dot{\gamma}}{}^\beta \tilde{S}_{\beta \alpha}
	- \frac{1}{4} \theta^2 \bar{\theta}^2 \Box \chi_\alpha \, , \\
Y_\alpha
	&= \psi_\alpha
	- \bar{\theta}^{\dot{\beta}} \tilde{R}_{\dot{\beta} \alpha}
	+ \bar{\theta}^2 \left( \tilde{\lambda}_\alpha - \frac{i}{2} \tilde{\omega}_\alpha \right)
	+ i \theta \dslash{\partial} \bar{\theta} \psi_\alpha
	+ \frac{i}{2} \theta^\gamma \bar{\theta}^2 \dslash{\partial}_\gamma{}^{\dot{\beta}} \tilde{R}_{\dot{\beta} \alpha}
	- \frac{1}{4} \theta^2 \bar{\theta}^2 \Box \psi_\alpha \, .
\end{align}
This can be used to calculate the contributions to the Lagrangian outlined in equation (\ref{Lcompact})
\begin{align}
\left( X^\alpha X_\alpha \right)_F
	&= \chi \tilde{\lambda}
	+ \frac{i}{2} \chi \tilde{\omega}
	- \frac{1}{2} \mathrm{Tr}{\left( \tilde{S}^T \tilde{S} \right)}
	+ \tilde{\lambda} \chi
	+ \frac{i}{2} \tilde{\omega} \chi \, , \displaybreak[3] \\
\left( Y^\alpha Y_\alpha \right)_F
	&= \psi \tilde{\lambda}
	- \frac{i}{2} \psi \tilde{\omega}
	- \frac{1}{2} \mathrm{Tr}{\left( \tilde{R}^T \tilde{R} \right)}
	+ \tilde{\lambda} \psi
	- \frac{i}{2} \tilde{\omega} \psi \, , \displaybreak[3] \\
\left( X^\alpha Y_\alpha \right)_D
	&= \partial_\mu \chi \partial^\mu \psi
	+ \tilde{\lambda} \tilde{\lambda}
	- \frac{i}{2} \tilde{\lambda} \tilde{\omega}
	+ \frac{i}{2} \tilde{\omega} \tilde{\lambda}
	+ \frac{1}{4} \tilde{\omega} \tilde{\omega}
	+ \frac{i}{2} \mathrm{Tr}{\left( \tilde{S}^T \dslash{\partial} \tilde{R} \right)} \, , \\
\left( Y^\alpha X_\alpha \right)_D
	&= \partial_\mu \psi \partial^\mu \chi
	+ \tilde{\lambda} \tilde{\lambda}
	+ \frac{i}{2} \tilde{\lambda} \tilde{\omega}
	- \frac{i}{2} \tilde{\omega} \tilde{\lambda}
	+ \frac{1}{4} \tilde{\omega} \tilde{\omega}
	+ \frac{i}{2} \mathrm{Tr}{\left( \tilde{R}^T \bar{\dslash{\partial}} \tilde{S} \right)} \, .
\end{align}
Therefore, the Lagrangian is given by
\begin{align}
\mathcal{L}
	&= \partial_\mu \chi \partial^\mu \psi
	+ \partial_\mu \psi \partial^\mu \chi
	+ 2 \tilde{\lambda} \tilde{\lambda}
	+ \frac{1}{2} \tilde{\omega} \tilde{\omega}
	+ \frac{m}{2} \chi \tilde{\lambda}
	+ \frac{i m}{4} \chi \tilde{\omega}
	+ \frac{m}{2} \tilde{\lambda} \chi
	+ \frac{i m}{4} \tilde{\omega} \chi + \notag \\
	&\quad + \frac{m}{2} \psi \tilde{\lambda}
	- \frac{i m}{4} \psi \tilde{\omega}
	+ \frac{m}{2} \tilde{\lambda} \psi
	- \frac{i m}{4} \tilde{\omega} \psi
	+ \frac{i}{2} \mathrm{Tr}{\left( \tilde{S}^T \dslash{\partial} \tilde{R} \right)}
	+ \frac{i}{2} \mathrm{Tr}{\left( \tilde{R}^T \bar{\dslash{\partial}} \tilde{S} \right)} - \notag \\
	&\quad - \frac{m}{4} \mathrm{Tr}{\left( \tilde{S}^T \tilde{S} \right)}
	- \frac{m}{4} \mathrm{Tr}{\left( \tilde{R}^T \tilde{R} \right)} 
	+ h.c. \, .
\label{Lmdimone}
\end{align}
It can be seen that this Lagrangian still contains the auxiliary fields $\tilde{\lambda}$ and $\tilde{\omega}$. They can be eliminated from the Lagrangian using their equations of motion
\begin{align}
\tilde{\omega}_\tau
	&= - \frac{i m}{2} \left( \chi_\tau - \psi_\tau \right) \, , \label{Mdimoneeqmomega} \\
\tilde{\lambda}_\tau
	&= - \frac{m}{4} \left( \chi_\tau + \psi_\tau \right)\, . \label{Mdimoneeqmlambda}
\end{align}
This process is also referred to as going "on-shell". The resulting on-shell Lagrangian is then found to be
\begin{align}
\mathcal{L}
	&= \partial_\mu \chi \partial^\mu \psi
	+ \partial_\mu \psi \partial^\mu \chi
	- \frac{m^2}{4} \psi \chi
	- \frac{m^2}{4} \chi \psi + \notag \\
	&\quad + \frac{i}{2} \mathrm{Tr}{\left( \tilde{S}^T \dslash{\partial} \tilde{R} \right)}
	+ \frac{i}{2} \mathrm{Tr}{\left( \tilde{R}^T \bar{\dslash{\partial}} \tilde{S} \right)}
	- \frac{m}{4} \mathrm{Tr}{\left( \tilde{S}^T \tilde{S} \right)}
	- \frac{m}{4} \mathrm{Tr}{\left( \tilde{R}^T \tilde{R} \right)} \, .
\label{Lonshell}
\end{align}
It is solely dependent on the on-shell component fields $\chi$, $\psi$, $\tilde{S}$, and $\tilde{R}$. On the first glance it seems that there are twice as many bosonic degrees of freedom as fermionic ones, because each of the second-rank spinor fields has in general 8 degrees of freedom while each of the complex spinor fields only encompasses four degrees of freedom. However, on-shell the bosonic second-rank spinor fields satisfy a Weyl type equation which reduces the number of bosonic on-shell degrees of freedom by a factor of 2. This means that the Lagrangian indeed has 8 fermionic and 8 bosonic degrees of freedom.

\section{The Supercurrent}
\label{CJonshell}
In classical field theory the Noether theorem describes the connection between symmetry transformations that leave the Lagrangian invariant and the corresponding conserved quantities. It states that every symmetry results in a conserved current which can alternatively be expressed as a conserved charge. Even though supersymmetry is not a symmetry in the classical sense the on-shell Lagrangian is invariant under the variation of the component fields as defined in equations (\ref{deltapsionshell}) to (\ref{deltaSonshell}). Therefore, according to Noether's theorem, a conserved supercurrent exists.

The general equation for the supercurrent is given by
\begin{align}
J_{\mu \kappa}
	&= \frac{\partial}{\partial \zeta^\kappa} \left( \sum\limits_\phi \delta \phi \frac{\partial \mathcal{L}}{\partial \partial^\mu \phi}
	- \mathcal{K}_\mu \right) \, ,
\end{align}
where the summation runs over all component fields of the Lagrangian. It has to be emphasised that this compact general equation for the supercurrent suppresses any indices of the component fields and also includes all hermitian conjugate component fields as well. The term $\mathcal{K}_\mu$ in this equation is related to the variation of the Lagrangian by
\begin{align}
\partial^\mu \mathcal{K}_\mu
	&= \delta \mathcal{L} \, .
\end{align}

As the full supercurrent $J_\mu$ is hermitian it is possible to restrict the discussion to the on-shell Lagrangian without hermitian conjugate part and to calculate both $J_\mu^{1/2}$ as well as $\bar{J}_\mu^{1/2}$. The complete supercurrent can then be constructed from the two contributions $J_\mu^{1/2}$ and $\bar{J}_\mu^{1/2}$.

The general equation for $J^{1/2}_\mu$ can be written as
\begin{align}
J^{1/2}_{\mu \kappa}
	&= \frac{\partial}{\partial \zeta^\kappa} \left( \delta \chi^\tau \frac{\partial \mathcal{L}}{\partial \partial^\mu \chi^\tau}
	+ \delta \psi^\tau \frac{\partial \mathcal{L}}{\partial \partial^\mu \psi^\tau}
	+ \delta \tilde{S}^{\tau \omega} \frac{\partial \mathcal{L}}{\partial \partial^\mu \tilde{S}^{\tau \omega}}
	+ \delta \tilde{R}^{\dot{\tau} \omega} \frac{\partial \mathcal{L}}{\partial \partial^\mu \tilde{R}^{\dot{\tau} \omega}}
	- \mathcal{K}_\mu \right) \, .
\end{align}
Inserting the on-shell Lagrangian from equation (\ref{Lonshell}) into the equation for the supercurrent yields
\begin{align}
J_{\mu \kappa}
	&= - 3 \tilde{S}_\kappa{}^\alpha \partial_\mu \psi_\alpha
	- \frac{i m}{2} \psi^\alpha \left( \sigma_\mu \right)_\kappa{}^{\dot{\beta}} \tilde{R}_{\dot{\beta} \alpha}
	- i \partial_\nu \psi^\alpha \left( \sigma^\nu{}_\mu \right)_\kappa{}^\beta \tilde{S}_{\beta \alpha}
	- \frac{\partial}{\partial \zeta^\kappa} \mathcal{K}_\mu \, .
\end{align}
The explicit form of $\mathcal{K}_\mu$ is derived from the variation of the Lagrangian without hermitian conjugate part
\begin{align}
\delta \mathcal{L}
	&= \partial_\mu \delta \chi \partial^\mu \psi
	+ \partial_\mu \chi \partial^\mu \delta \psi
	+ \partial_\mu \delta \psi \partial^\mu \chi
	+ \partial_\mu \psi \partial^\mu \delta \chi
	- \frac{m^2}{4} \delta \psi \chi
	- \frac{m^2}{4} \psi \delta \chi
	- \frac{m^2}{4} \delta \chi \psi - \notag \\
	&\quad - \frac{m^2}{4} \chi \psi \delta + \frac{i}{2} \mathrm{Tr}{\left( \delta \tilde{S}^T \dslash{\partial} \tilde{R} \right)}
	+ \frac{i}{2} \mathrm{Tr}{\left( \tilde{S}^T \dslash{\partial} \delta \tilde{R} \right)}
	+ \frac{i}{2} \mathrm{Tr}{\left( \delta \tilde{R}^T \bar{\dslash{\partial}} \tilde{S} \right)}
	+ \frac{i}{2} \mathrm{Tr}{\left( \tilde{R}^T \bar{\dslash{\partial}} \delta \tilde{S} \right)} - \notag \\
	&\quad - \frac{m}{4} \mathrm{Tr}{\left( \delta \tilde{S}^T \tilde{S} \right)}
	- \frac{m}{4} \mathrm{Tr}{\left( \tilde{S}^T \delta \tilde{S} \right)}
	- \frac{m}{4} \mathrm{Tr}{\left( \delta \tilde{R}^T \tilde{R} \right)}
	- \frac{m}{4} \mathrm{Tr}{\left( \tilde{R}^T \delta \tilde{R} \right)} \, .
\end{align}	
It can be shown that the variation of the Lagrangian is a four divergence as expected which implies that
\begin{align}
\mathcal{K}_\mu
	&= \zeta^\beta \tilde{S}_{\beta \alpha} \partial_\mu \psi^\alpha
	- \bar{\zeta}^{\dot{\beta}} \tilde{R}_{\dot{\beta} \alpha} \partial_\mu \chi^\alpha
	+ \frac{i m}{2} \left( \bar{\sigma}_\mu \right)_{\dot{\beta}}{}^\gamma \bar{\zeta}^{\dot{\beta}} \chi^\alpha  \tilde{S}_{\gamma \alpha}
	+ \frac{i m}{2} \left( \sigma_\mu \right)_\beta{}^{\dot{\gamma}} \zeta^\beta \psi^\alpha \tilde{R}_{\dot{\gamma} \alpha} + \notag \\
	&\quad + i \bar{\zeta}^{\dot{\delta}} \left( \bar{\sigma}_\mu{}^\nu \right)_{\dot{\delta}}{}^{\dot{\gamma}} \chi^\alpha \partial_\nu \tilde{R}_{\dot{\gamma} \alpha}
	+ i \zeta^\delta \left( \sigma_\mu{}^\nu \right)_\delta{}^\gamma \psi^\alpha \partial_\nu \tilde{S}_{\gamma \alpha} \, .
\label{Kmu}
\end{align}
This result can then be inserted into the equation for the supercurrent. After differentiating with respect to the transformation parameter $\zeta$ the supercurrent is found to be
\begin{align}
J^{1/2}_{\mu \kappa}
	&= - i m \left( \sigma_\mu \right)_\kappa{}^{\dot{\beta}} \tilde{R}_{\dot{\beta} \alpha} \psi^\alpha
	+ 2 \left( \sigma_\mu \right)^{\beta \dot{\gamma}} \bar{\dslash{\partial}}_{\dot{\gamma} \kappa} \psi^\alpha \tilde{S}_{\beta \alpha} \, .
\label{Jhalf}
\end{align}

The contribution to the full supercurrent $\bar{J}_\mu^{1/2}$ is defined in perfect analogy to $J_\mu^{1/2}$ by replacing the derivative with respect to the Grassmann variable $\zeta$ with a derivative with respect to $\bar{\zeta}$. It has to be noted that the behaviour of the Grassmann derivative is rather subtle and depends on the conventions chosen. In the present scenario where by convention all derivatives are written as right derivatives the change from left to right derivative introduces an additional overall minus sign
\begin{align}
\bar{J}^{1/2}_{\mu \dot{\kappa}}
	&= - \frac{\partial}{\partial \bar{\zeta}^{\dot{\kappa}}} \left( \delta \chi^\tau \frac{\partial \mathcal{L}}{\partial \partial^\mu \chi^\tau}
	+ \delta \psi^\tau \frac{\partial \mathcal{L}}{\partial \partial^\mu \psi^\tau}
	+ \delta \tilde{S}^{\tau \omega} \frac{\partial \mathcal{L}}{\partial \partial^\mu \tilde{S}^{\tau \omega}}
	+ \delta \tilde{R}^{\dot{\tau} \omega} \frac{\partial \mathcal{L}}{\partial \partial^\mu \tilde{R}^{\dot{\tau} \omega}}
	- \mathcal{K}_\mu \right) \, .
\end{align}
The supercurrent $\bar{J}_\mu^{1/2}$ for the Lagrangian without the complex conjugate part is then given by
\begin{align}
\bar{J}^{1/2}_{\mu \dot{\kappa}}
	&= 3 \tilde{R}_{\dot{\kappa} \alpha} \partial_\mu \chi^\alpha
	+ i \left( \sigma^\nu{}_\mu \right)_{\dot{\kappa}}{}^{\dot{\beta}} \partial_\nu \chi^\alpha \tilde{R}_{\dot{\beta} \alpha}
	+ \frac{i m}{2} \left( \bar{\sigma}_\mu \right)_{\dot{\kappa}}{}^\beta \tilde{S}_{\beta \alpha} \chi^\alpha
	+ \frac{\partial}{\partial \bar{\zeta}^{\dot{\kappa}}} \mathcal{K}_\mu \, ,
\end{align}
where the term $\mathcal{K}_\mu$ is already known from equation (\ref{Kmu}). After differentiation with respect to the Grassmann variable the final result is
\begin{align}
\bar{J}^{1/2}_{\mu \dot{\kappa}}
	&= i m \left( \bar{\sigma}_\mu \right)_{\dot{\kappa}}{}^\beta \tilde{S}_{\beta \alpha} \chi^\alpha
	+ 2 \left( \bar{\sigma}_\mu \right)^{\dot{\beta} \gamma} \dslash{\partial}_{\gamma \dot{\kappa}} \chi^\alpha \tilde{R}_{\dot{\beta} \alpha} \, .
\label{Jbarhalf}
\end{align}
Together with the previous result for $J^{1/2}_\mu$ from equation (\ref{Jhalf}) the construction of the full supercurrent is straightforward
\begin{align}
J_{\mu \kappa}
	&= - i m \left( \sigma_\mu \right)_\kappa{}^{\dot{\beta}} \tilde{R}_{\dot{\beta} \alpha} \psi^\alpha
	+ 2 \left( \sigma_\mu \right)^{\beta \dot{\gamma}} \bar{\dslash{\partial}}_{\dot{\gamma} \kappa} \psi^\alpha \tilde{S}_{\beta \alpha} - \notag \\
	&\quad - i m \left( \sigma_\mu \right)_{\kappa}{}^{\dot{\beta}} \bar{\tilde{S}}_{\dot{\beta} \dot{\alpha}} \bar{\chi}^{\dot{\alpha}}
	+ 2 \left( \sigma_\mu \right)^{\beta \dot{\gamma}} \bar{\dslash{\partial}}_{\dot{\gamma} \kappa} \bar{\chi}^{\dot{\alpha}} \bar{\tilde{R}}_{\beta \dot{\alpha}} \, .
\label{Jfull}
\end{align}

\section{The Hamiltonian in Position Space}
\label{CHposition}
The Hamiltonian in position space is usually derived from the Lagrangian by canonical quantisation. However, it is not immediately clear whether this approach is still valid for the present scenario that is based on a general spinor superfield instead of a scalar superfield. Therefore, a more conservative approach based on the supersymmetry algebra was chosen. It utilises the anticommutation relation between the barred and unbarred supersymmetry generator of the $N = 1$ supersymmetry algebra
\begin{align}
2 \left( \sigma^\mu \right)_{\alpha \dot{\beta}} P_\mu
	&= 
	\left\{ Q_\alpha , \bar{Q}_{\dot{\beta}} \right\} .
\label{PSUSYalgebra}
\end{align}

At this point it can already be seen that a successful derivation of the Hamiltonian from the supersymmetry algebra requires the knowledge of the commutation and anticommutation relations of the component fields in position space. Therefore, the second quantisation of the component fields in position space will be discussed in Section \ref{SQuantCompFields}. Afterwards in Section \ref{CHSUSYalgebra}, these results will be used to derive an expression for the Hamiltonian in position space which is founded in the supersymmetry algebra. Finally, in Section \ref{CHcanonicalquant} it will be shown that canonical quantisation yields the same Hamiltonian in position space as the approach based on the supersymmetry algebra.

\subsection{Second Quantisation in Position Space}
\label{SQuantCompFields}
A viable supersymmetric model of fermionic fields with mass dimension one requires a second quantisation that is in agreement with the superfield transformations of the component fields as derived in Section \ref{SSSUSYtranslation}. This can be achieved by calculating the commutator between the component fields and the generators of the superspace translations
\begin{align}
\delta \phi
	&= - i \left[ \phi , \zeta^\alpha Q_\alpha + \zeta_{\dot{\alpha}} Q^{\dot{\alpha}} \right] \, .
\label{compfieldvarcommutator}
\end{align}
To generalise the notation the spinor indices of the field $\phi$ are suppressed and it can represent a scalar field as well as first, second, or higher rank spinor fields. Subsequently, the commutation and anticommutation relations of the barred component fields are derived from the results for the unbarred component fields.

The supersymmetry generators that appear in this equation are proportional to the supercurrent
\begin{align}
Q_\alpha
	&= \int \mathrm{d} \mathbf{x} J_{0 \alpha} \, , \label{QproptoJ}\\
\bar{Q}_{\dot{\alpha}} 
	&= \int \mathrm{d} \mathbf{x} \bar{J}_{0 \dot{\alpha}} \, . \label{QbarproptoJbar}
\end{align}
In general the supersymmetry generators must contain the full supercurrent. However, the previous results for the superfield translations, equations (\ref{deltapsionshell}) to (\ref{deltaSonshell}), imply that no mixing between barred and unbarred component fields occurs. Therefore, it is sufficient to restrict the discussion in this section to the supercurrent arising from the Lagrangian without hermitian conjugate contribution, as any cross terms vanish identically and define the constrained generators
\begin{align}
Q^{1/2}_\alpha
	&= \int \mathrm{d} \mathbf{x} J^{1/2}_{0 \alpha} \, , \\
\bar{Q}^{1/2}_{\dot{\alpha}} 
	&= \int \mathrm{d} \mathbf{x} \bar{J}^{1/2}_{0 \dot{\alpha}} \, .
\end{align}
To distinguish the constrained generators from the full generators as outlined in equations (\ref{QproptoJ}) and (\ref{QbarproptoJbar}) an additional superscript $1/2$ was incorporated into the notation in analogy to the notation for the supercurrent in Chapter \ref{CJonshell}. Inserting the results for the supercurrent from equations (\ref{Jhalf}) and (\ref{Jbarhalf}) then yields the following expression for the constrained supersymmetry generators
\begin{align}
Q^{1/2}_\alpha
	&= \int \mathrm{d} \mathbf{x} \left( - i m \left( \sigma_\mu \right)_\alpha{}^{\dot{\gamma}} \tilde{R}_{\dot{\gamma} \beta}{\left( x \right)} \psi^\beta{\left( x \right)}
	+ 2 \left( \sigma_\mu \right)^{\gamma \dot{\delta}} \tilde{S}_{\gamma \beta}{\left( x \right)} \bar{\dslash{\partial}}_{\dot{\delta} \alpha} \psi^\beta{\left( x \right)} \right) \, , \label{generatorQ}\\
\bar{Q}^{1/2}_{\dot{\alpha}}
	&= \int \mathrm{d} \mathbf{x} \left( i m \left( \bar{\sigma}_\mu \right)_{\dot{\alpha}}{}^\gamma \tilde{S}_{\gamma \beta}{\left( x \right)} \chi^\beta{\left( x \right)}
	+ 2 \left( \bar{\sigma}_\mu \right)^{\dot{\gamma} \delta} \tilde{R}_{\dot{\gamma} \beta}{\left( x \right)} \dslash{\partial}_{\delta \dot{\alpha}} \chi^\beta{\left( x \right)} \right) \, . \label{generatorQbar}
\end{align}

\subsubsection{Superfield Transformation of the Fermionic Component Fields}
Inserting the constrained supersymmetry generators as defined in equations (\ref{generatorQ}) and (\ref{generatorQbar}) into equation (\ref{compfieldvarcommutator}) for the commutator between component field $\chi$ and the generators of superspace translations yields a variation of $\chi$ of
\begin{align}
\delta \chi_\alpha{\left( x \right)}
	&= \int \mathrm{d} \mathbf{x}' \left( m \zeta^\beta \left( \sigma_0 \right)_\beta{}^{\dot{\gamma}} \left\{ \chi_\alpha{\left( x \right)} , \tilde{R}_{\dot{\gamma} \delta}{\left( x' \right)} \psi^\delta{\left( x' \right)} \right\} + \right. \notag \\
	&\qquad + 2 i \zeta^\beta \left( \sigma_0 \right)^{\gamma \dot{\delta}} \left\{ \chi_\alpha{\left( x \right)} , \tilde{S}_{\gamma \epsilon}{\left( x' \right)} \bar{\dslash{\partial}}'_{\dot{\delta} \beta} \psi^\epsilon{\left( x' \right)} \right\} - \notag \\
	&\qquad - m \bar{\zeta}_{\dot{\beta}} \left( \bar{\sigma}_0 \right)^{\dot{\beta} \gamma} \left\{ \chi_\alpha{\left( x \right)} , \tilde{S}_{\gamma \delta}{\left( x' \right)} \chi^\delta{\left( x' \right)} \right\} + \notag \\
	&\qquad \left. + 2 i \bar{\zeta}_{\dot{\beta}} \left( \bar{\sigma}_0 \right)^{\dot{\gamma} \delta} \left\{ \chi_\alpha{\left( x \right)} , \tilde{R}_{\dot{\gamma} \epsilon}{\left( x' \right)} \dslash{\partial}'_\delta{}^{\dot{\beta}} \chi^\epsilon{\left( x' \right)} \right\} \right) \, .
\end{align}
Each of the contributions to the variation of the component field $\chi$ contains an anticommutator involving two fermionic fields and one bosonic field. They can be rewritten using the anticommutator relation
\begin{align}
\left\{ F_1 , B_2 F_2 \right\}
	&= B_2 \left\{ F_1 , F_2 \right\} \, .
\end{align}
In addition, it can be seen that the second and fourth term contain a four derivative $\dslash{\partial}$ acting on one of the component fields in the anticommutator. These terms can be rewritten by splitting the four derivative into its time and spatial components
\begin{align}
\dslash{\partial}_{\alpha \dot{\beta}}
	&= \left( \sigma^\mu \right)_{\alpha \dot{\beta}} \partial_\mu
	= \left( \sigma^0 \right)_{\alpha \dot{\beta}} \partial_0
	+ \boldsymbol{\sigma}_{\alpha \dot{\beta}} \cdot \boldsymbol{\nabla} \, .
\end{align}
It is important to recall that there is a plus sign between the time and spatial components and not a minus sign as the derivative is a covariant three vector $\boldsymbol{\nabla} = \left( \partial_1 , \partial_2 , \partial_3 \right)$ while all standard vectors, e.g. $\mathbf{p} = \left( p^1, p^2 , p^3 \right)$, are contravariant three vectors. After partial integration over the spatial components each term involving a four-derivative is replaced by two terms -- one containing a time derivative acting on one of the fields in the commutator and one simply containing the commutator of component fields. Furthermore, the boundary terms from the partial integration which are 3-divergences vanish identically and are ignored. This results in
\begin{align}
\delta \chi_\alpha{\left( x \right)}
	&= \int \mathrm{d} \mathbf{x}' \left( m \zeta^\beta \left( \sigma_0 \right)_\beta{}^{\dot{\gamma}} \tilde{R}_{\dot{\gamma} \delta}{\left( x' \right)} \left\{ \chi_\alpha{\left( x \right)} , \psi^\delta{\left( x' \right)} \right\}
	+ 2 i \zeta^\beta \tilde{S}_{\beta \epsilon}{\left( x' \right)} \left\{ \chi_\alpha{\left( x \right)} , \dot{\psi}^\epsilon{\left( x' \right)} \right\} + \right. \notag \\
	&\qquad + 2 i \zeta^\beta \left( \sigma_0 \right)^{\gamma \dot{\delta}} \boldsymbol{\sigma}_{\dot{\delta} \beta} \cdot \boldsymbol{\nabla}' \tilde{S}_{\gamma \epsilon}{\left( x' \right)} \left\{ \chi_\alpha{\left( x \right)} , \psi^\epsilon{\left( x' \right)} \right\} - \notag \\
	&\qquad - m \bar{\zeta}_{\dot{\beta}} \left( \bar{\sigma}_0 \right)^{\dot{\beta} \gamma} \tilde{S}_{\gamma \delta}{\left( x' \right)} \left\{ \chi_\alpha{\left( x \right)} , \chi^\delta{\left( x' \right)} \right\} 
	- 2 i \bar{\zeta}^{\dot{\beta}} \tilde{R}_{\dot{\beta} \epsilon}{\left( x' \right)} \left\{ \chi_\alpha{\left( x \right)} , \dot{\chi}^\epsilon{\left( x' \right)} \right\} + \notag \\
	&\qquad \left. + 2 i \bar{\zeta}_{\dot{\beta}} \left( \bar{\sigma}_0 \right)^{\dot{\gamma} \delta} \boldsymbol{\sigma}_\delta{}^{\dot{\beta}} \cdot  \boldsymbol{\nabla}' \tilde{R}_{\dot{\gamma} \epsilon}{\left( x' \right)} \left\{ \chi_\alpha{\left( x \right)} , \chi^\epsilon{\left( x' \right)} \right\} \right) \, .
\label{deltachisecondquant}
\end{align}
By comparison with the previously derived superspace translation of $\chi$ in equation (\ref{deltachionshell}) it can be seen that the only nonvanishing contribution comes from the term proportional to $\zeta \tilde{S}$ while all other contributions have to vanish identically. This implies that three of the anticommutators vanish identically
\begin{align}
\left\{ \chi_\alpha{\left( x \right)} , \psi_\beta{\left( x' \right)} \right\}
	&= 0 \, ,\\
\left\{ \chi_\alpha{\left( x \right)} , \dot{\chi}_\beta{\left( x' \right)} \right\}
	&= 0 \, , \\
\left\{ \chi_\alpha{\left( x \right)} , \chi_\beta{\left( x' \right)} \right\}
	&= 0 \, .
\end{align}
The only nonvanishing anticommutator satisfies
\begin{align}
- \zeta^\beta \tilde{S}_{\beta \alpha}{\left( x \right)}
	&= - 2 i \zeta^\beta \int \mathrm{d} \mathbf{x}' \tilde{S}_\beta{}^\gamma{\left( x' \right)} \left\{ \chi_\alpha{\left( x \right)} , \dot{\psi}_\gamma{\left( x' \right)} \right\} \, ,
\label{quantrelchi}
\end{align}
which has the solution
\begin{align}
\left\{ \chi_\alpha{\left( x \right)} , \dot{\psi}_\gamma{\left( x' \right)} \right\}
	&= \frac{i}{2} \epsilon_{\alpha \gamma} \delta{\left( x - x' \right)} \, .
\end{align}

As the Lagrangian is symmetric with respect to the exchange of $\chi$ and $\psi$ there is no difference between the calculation of $\delta \chi$ and $\delta \psi$. Again, three of the anticommutators have to vanish identically
\begin{align}
\left\{ \psi_\alpha{\left( x \right)} , \dot{\psi}_\beta{\left( x' \right)} \right\}
	&= 0 \, , \\
\left\{ \psi_\alpha{\left( x \right)} , \psi_\beta{\left( x' \right)} \right\}
	&= 0 \, , \\
\left\{ \psi_\alpha{\left( x \right)} , \chi_\beta{\left( x' \right)} \right\}
	&= 0\, ,
\end{align}
while the only nonvanishing anticommutator is the one involving $\psi$ and $\dot{\chi}$
\begin{align}
\left\{ \psi_\alpha{\left( x \right)} , \dot{\chi}_\gamma{\left( x' \right)} \right\}
	&= \frac{i}{2} \epsilon_{\alpha \gamma} \delta{\left( \mathbf{x} - \mathbf{x'} \right)} \, .
\end{align}

\subsubsection{Superfield Transformation of the Bosonic Component Fields}
The discussion for the superfield transformation of the bosonic component fields is in perfect analogy to those for the fermionic component fields. The change from a fermionic to a bosonic field results in an exchange of all anticommutators with commutators
\begin{align}
\delta \tilde{S}_{\beta \alpha}{\left( x \right)}
	&= \int \mathrm{d} \mathbf{x}' \left( - m \zeta^\gamma \left( \sigma_0 \right)_\gamma{}^{\dot{\epsilon}} \left[ \tilde{S}_{\beta \alpha}{\left( x \right)} , \tilde{R}_{\dot{\epsilon} \delta}{\left( x' \right)} \psi^\delta{\left( x' \right)} \right] - \right. \notag \\
	&\qquad- 2 i \zeta^\gamma \left( \sigma_0 \right)^{\epsilon \dot{\delta}} \left[ \tilde{S}_{\beta \alpha}{\left( x \right)} , \tilde{S}_{\epsilon \kappa}{\left( x' \right)} \bar{\dslash{\partial}}_{\dot{\delta} \gamma} \psi^\kappa{\left( x' \right)} \right] +  \notag \\
	&\qquad + m \bar{\zeta}_{\dot{\gamma}} \left( \bar{\sigma}_0 \right)^{\dot{\gamma} \epsilon} \left[ \tilde{S}_{\beta \alpha}{\left( x \right)} , \tilde{S}_{\epsilon \delta}{\left( x' \right)} \chi^\delta{\left( x' \right)} \right] - \notag \\
	&\qquad \left. - 2 i \bar{\zeta}_{\dot{\gamma}} \left( \bar{\sigma}_0 \right)^{\dot{\epsilon} \delta} \left[ \tilde{S}_{\beta \alpha}{\left( x \right)} , \tilde{R}_{\dot{\epsilon} \kappa}{\left( x' \right)} \dslash{\partial}_\delta{}^{\dot{\gamma}} \chi^\kappa{\left( x' \right)} \right] \right) \, .
\end{align}
The commutators involved in this expression each contain two bosonic and one fermionic component field and can be simplified using the commutator relation
\begin{align}
\left[ B_1 , B_2 F_2 \right]
	&= F_2 \left[ B_1 , B_2 \right] \, .
\end{align}
Therefore, the variation of the bosonic second-rank spinor field $\tilde{S}$ takes the form
\begin{align}
\delta \tilde{S}_{\beta \alpha}{\left( x \right)}
	&= \int \mathrm{d} \mathbf{x}' \left( - m \zeta^\gamma \left( \sigma_0 \right)_\gamma{}^{\dot{\epsilon}} \psi^\delta{\left( x' \right)} \left[ \tilde{S}_{\beta \alpha}{\left( x \right)} , \tilde{R}_{\dot{\epsilon} \delta}{\left( x' \right)} \right] - \right. \notag \\
	&\qquad - 2 i \zeta^\gamma \left( \sigma_0 \right)^{\epsilon \dot{\delta}} \bar{\dslash{\partial}}_{\dot{\delta} \gamma} \psi^\kappa{\left( x' \right)} \left[ \tilde{S}_{\beta \alpha}{\left( x \right)} , \tilde{S}_{\epsilon \kappa}{\left( x' \right)} \right] + \notag \\
	&\qquad + m \bar{\zeta}_{\dot{\gamma}} \left( \bar{\sigma}_0 \right)^{\dot{\gamma} \epsilon} \chi^\delta{\left( x' \right)} \left[ \tilde{S}_{\beta \alpha}{\left( x \right)} , \tilde{S}_{\epsilon \delta}{\left( x' \right)} \right] - \notag \\
	&\qquad \left. - 2 i \bar{\zeta}_{\dot{\gamma}} \left( \bar{\sigma}_0 \right)^{\dot{\epsilon} \delta} \dslash{\partial}_\delta{}^{\dot{\gamma}} \chi^\kappa{\left( x' \right)} \left[ \tilde{S}_{\beta \alpha}{\left( x \right)} , \tilde{R}_{\dot{\epsilon} \kappa}{\left( x' \right)} \right] \right) \, .
\label{deltaSsecondquant}
\end{align}
If this result is compared to the superspace translation of $\tilde{S}$ in equation (\ref{deltaSonshell}) it can be immediately read off that
\begin{align}
\left[ \tilde{S}_{\beta \alpha}{\left( x \right)} , \tilde{S}_{\gamma \delta}{\left( x' \right)} \right]
	&= 0 \, .
\end{align}
The remaining two relations for the commutator between $\tilde{S}$ and $\tilde{R}$ should yield the same result. Using the first relation 
\begin{align}
m \zeta_\beta \psi_\alpha
	&= - m \zeta^\gamma \left( \sigma_0 \right)_\gamma{}^{\dot{\epsilon}} \int \mathrm{d} \mathbf{x}' \psi^\delta{\left( x' \right)} \left[ \tilde{S}_{\beta \alpha}{\left( x \right)} , \tilde{R}_{\dot{\epsilon} \delta}{\left( x' \right)} \right]
\label{quantrelS}
\end{align}
it is found that $\tilde{S}$ and $\tilde{R}$ satisfy the commutation relation
\begin{align}
\left[ \tilde{S}_{\beta \alpha}{\left( x \right)} , \tilde{R}_{\dot{\epsilon} \delta}{\left( x' \right)} \right]
	&= - \epsilon_{\alpha \delta} \delta{\left( x - x' \right)}\left( \sigma^0 \right)_{\beta \dot{\epsilon}} \, .
\label{commutatorSR}
\end{align}
Inserting this result into the second relation then provides a consistency check as it satisfies the relation identically.

Again, the calculations for the superfield transformation of $\tilde{R}$ are in perfect analogy to those for $\tilde{S}$. It is found that the commutator of $\tilde{R}$ with itself vanishes
\begin{align}
\left[ \tilde{R}_{\dot{\beta} \alpha}{\left( x \right)} , \tilde{R}_{\dot{\gamma} \delta}{\left( x' \right)} \right]
	&= 0 \, .
\end{align}
Finally, the remaining commutator between $\tilde{R}$ and $\tilde{S}$ can be derived from equation (\ref{commutatorSR}) by commuting the component fields and renaming the spinor indices appropriately
\begin{align}
\left[ R_{\dot{\beta} \alpha}{\left( x \right)} , S'_{\epsilon \delta}{\left( x' \right)} \right]
	&= - \epsilon_{\alpha \delta} \left( \bar{\sigma}^0 \right)_{\dot{\beta} \epsilon} \delta{\left( x - x' \right)} \, .
\label{commutatorRS}
\end{align}

\subsubsection{Transformation of the Conjugate Component Fields}
Generally it is possible to repeat the calculations outlined in the previous sections for the hermitian conjugate component fields. However, it is much easier to calculate the hermitian conjugate of the previously derived commutation and anticommutation relations.

For the anticommutation relations between the spinor fields hermitian conjugation is straightforward and the two nonvanishing anticommutation relations between the barred component fields are
\begin{align}
\left\{ \bar{\chi}_{\dot{\alpha}}{\left( x \right)} , \dot{\bar{\psi}}_{\dot{\gamma}}{\left( x' \right)} \right\}
	&= \frac{i}{2} \epsilon_{\dot{\alpha} \dot{\gamma}} \delta{\left( \mathbf{x} - \mathbf{x}' \right)} \, , \\
\left\{ \bar{\psi}_{\dot{\alpha}}{\left( x \right)} , \dot{\bar{\chi}}_{\dot{\gamma}}{\left( x' \right)} \right\}
	&= \frac{i}{2} \epsilon_{\dot{\alpha} \dot{\gamma}} \delta{\left( \mathbf{x} - \mathbf{x'} \right)} \, .
\end{align}
The only difficulty that arises is the sign change of the second-rank $\epsilon$-tensor under hermitian conjugation.

For the commutation relations of the bosonic second-rank spinor fields the discussion is only slightly more involved as the hermitian conjugation inverts the ordering of the component fields which induces an additional sign flip for the commutators that didn't occur for the spinor fields. Therefore, the commutation relations for the barred component fields are given by
\begin{align}
\left[ \tilde{\bar{S}}_{\dot{\beta} \dot{\alpha}}{\left( x \right)} , \tilde{\bar{R}}_{\epsilon \dot{\delta}}{\left( x' \right)} \right]
	&= - \epsilon_{\dot{\alpha} \dot{\delta}} \delta{\left( \mathbf{x} - \mathbf{x}' \right)} \left( \bar{\sigma}^0 \right)_{\dot{\beta} \epsilon} \, , \\
\left[ \tilde{\bar{R}}_{\beta \dot{\alpha}}{\left( x \right)} , \tilde{\bar{S}}_{\dot{\epsilon} \dot{\delta}}{\left( x' \right)} \right]
	&= - \epsilon_{\dot{\alpha} \dot{\delta}} \left( \sigma^0 \right)_{\beta \dot{\epsilon}} \delta{\left( \mathbf{x} - \mathbf{x}' \right)} \, .
\end{align}

\subsection{The Hamiltonian from the Supersymmetry Algebra}
\label{CHSUSYalgebra}
To derive an explicit equation for the Hamiltonian the supersymmetry generators in equation (\ref{PSUSYalgebra}) have to be expressed in terms of the component fields. This can be achieved using the relations between the supersymmetry generators which are the conserved Noether charges of the system and the supercurrents which were defined in equations (\ref{QproptoJ}) and (\ref{QbarproptoJbar}). Inserting the result for the supercurrent from equations (\ref{Jfull}) and its hermitian conjugate leads to the following expression of the supersymmetry generators in terms of the component fields
\begin{align}
Q_\alpha
	&= \int \mathrm{d} \mathbf{x} \left( - i m \left( \sigma_0 \right)_\alpha{}^{\dot{\gamma}} \tilde{R}_{\dot{\gamma} \beta}{\left( x \right)} \psi^\beta{\left( x \right)}
	+ 2 \left( \sigma_0 \right)^{\gamma \dot{\delta}} \tilde{S}_{\gamma \beta}{\left( x \right)} \bar{\dslash{\partial}}_{\dot{\delta} \alpha} \psi^\beta{\left( x \right)} - \right. \notag \\
	&\qquad \left. - i m \left( \sigma_0 \right)_\alpha{}^{\dot{\gamma}} \tilde{\bar{S}}_{\dot{\gamma} \dot{\beta}}{\left( x \right)} \bar{\chi}^{\dot{\beta}}{\left( x \right)}
	+ 2 \left( \sigma_0 \right)^{\gamma \dot{\delta}} \tilde{\bar{R}}_{\gamma \dot{\beta}}{\left( x \right)} \bar{\dslash{\partial}}_{\dot{\delta} \alpha} \bar{\chi}^{\dot{\beta}}{\left( x \right)} \right) \, , \displaybreak[3] \\
\bar{Q}_{\dot{\alpha}}
	&= \int \mathrm{d} \mathbf{x} \left( i m \left( \bar{\sigma}_0 \right)_{\dot{\alpha}}{}^\gamma \tilde{S}_{\gamma \beta}{\left( x \right)} \chi^\beta{\left( x \right)}
	+ 2 \left( \bar{\sigma}_0 \right)^{\dot{\gamma} \delta} \tilde{R}_{\dot{\gamma} \beta}{\left( x \right)} \dslash{\partial}_{\delta \dot{\alpha}} \chi^\beta{\left( x \right)} + \right. \notag \\
	&\qquad \left. + i m \left( \bar{\sigma}_0 \right)_{\dot{\alpha}}{}^\gamma \tilde{\bar{R}}_{\gamma \dot{\beta}}{\left( x \right)} \bar{\psi}^{\dot{\beta}}{\left( x \right)}
	+ 2 \left( \bar{\sigma}_0 \right)^{\dot{\gamma} \delta} \tilde{\bar{S}}_{\dot{\gamma} \dot{\beta}}{\left( x \right)} \dslash{\partial}_{\delta \dot{\alpha}} \bar{\psi}^{\dot{\beta}}{\left( x \right)} \right) \, .
\end{align}
To streamline the notation it proves useful to introduce the short notation
\begin{align}
\dslash{P}_{\alpha \dot{\beta}}
	&= \left( \sigma^\mu \right)_{\alpha \dot{\beta}} P_\mu \, ,
\end{align}
which is defined in analogy to the commonly used contraction with Dirac matrices. The momentum operator is then given by
\begin{align}
2 \dslash{P}_{\alpha \dot{\beta}}
	&= \left\{ \int \mathrm{d} \mathbf{x} \left( - i m \left( \sigma_0 \right)_\alpha{}^{\dot{\gamma}} \tilde{R}_{\dot{\gamma} \omega}{\left( x \right)} \psi^\omega{\left( x \right)}
	+ 2 \left( \sigma_0 \right)^{\gamma \dot{\delta}} \tilde{S}_{\gamma \omega}{\left( x \right)} \bar{\dslash{\partial}}_{\dot{\delta} \alpha} \psi^\omega{\left( x \right)} - \right. \right. \notag \\
	&\qquad \left. - i m \left( \sigma_0 \right)_\alpha{}^{\dot{\gamma}} \tilde{\bar{S}}_{\dot{\gamma} \dot{\omega}}{\left( x \right)} \bar{\chi}^{\dot{\omega}}{\left( x \right)}
	+ 2 \left( \sigma_0 \right)^{\gamma \dot{\delta}} \tilde{\bar{R}}_{\gamma \dot{\omega}}{\left( x \right)} \bar{\dslash{\partial}}'_{\dot{\delta} \alpha} \bar{\chi}^{\dot{\omega}}{\left( x \right)} \right) , \notag \\
	&\quad \int \mathrm{d} \mathbf{x}' \left( i m \left( \bar{\sigma}_0 \right)_{\dot{\beta}}{}^\kappa \tilde{S}_{\kappa \epsilon}{\left( x' \right)} \chi^\epsilon{\left( x' \right)}
	+ 2 \left( \bar{\sigma}_0 \right)^{\dot{\kappa} \tau} \tilde{R}_{\dot{\kappa} \epsilon}{\left( x' \right)} \dslash{\partial}'_{\tau \dot{\beta}} \chi^\epsilon{\left( x' \right)} + \right. \notag \\
	&\qquad \left. \left. + i m \left( \bar{\sigma}_0 \right)_{\dot{\beta}}{}^\kappa \tilde{\bar{R}}_{\kappa \dot{\epsilon}}{\left( x' \right)} \bar{\psi}^{\dot{\epsilon}}{\left( x' \right)}
	+ 2 \left( \bar{\sigma}_0 \right)^{\dot{\kappa} \tau} \tilde{\bar{S}}_{\dot{\kappa} \dot{\epsilon}}{\left( x' \right)} \dslash{\partial}_{\tau \dot{\beta}} \bar{\psi}^{\dot{\epsilon}}{\left( x' \right)} \right) \right\} \, .
\end{align}
The anticommutators containing two fermionic and two bosonic component fields can now  be rewritten using the commutator relation
\begin{align}
\left\{ B_1 F_1 , B_2 F_2 \right\}
	&= \left[ B_1 , B_2 \right] F_1 F_2 + B_2 B_1 \left\{ F_1 , F_2 \right\} \, ,
\end{align}
where it was assumed that the fermionic and bosonic fields commute. This assumption is justified by the previous derivation of the commutation and anticommutation relations of the component fields as well as the results of the superfield translations.

After separation of the time and spatial derivatives as well as partial spatial integration the momentum operator is given by
\begin{align}
2 \dslash{P}_{\alpha \dot{\beta}}
	&= \int \mathrm{d} \mathbf{x} \mathrm{d} \mathbf{x}' \left(
	m^2 \left( \sigma_0 \right)_\alpha{}^{\dot{\gamma}} \left( \bar{\sigma}_0 \right)_{\dot{\beta}}{}^\kappa \psi^\omega{\left( x \right)} \chi^\epsilon{\left( x' \right)} \left[ \tilde{R}_{\dot{\gamma} \omega}{\left( x \right)} , \tilde{S}_{\kappa \epsilon}{\left( x' \right)} \right] - \right. \notag \\
	&\qquad - 2 i m \left( \sigma_0 \right)_\alpha{}^{\dot{\gamma}} \left( \bar{\sigma}_0 \right)^{\dot{\kappa} \tau} \left( \sigma^0 \right)_{\tau \dot{\beta}} \tilde{R}_{\dot{\kappa} \epsilon}{\left( x' \right)} \tilde{R}_{\dot{\gamma} \omega}{\left( x \right)} \left\{ \psi^\omega{\left( x \right)} , \dot{\chi}^\epsilon{\left( x' \right)} \right\} + \notag \displaybreak[3] \\
	&\qquad + 2 i m \left( \sigma_0 \right)^{\gamma \dot{\delta}} \left( \bar{\sigma}_0 \right)_{\dot{\beta}}{}^\kappa \left( \bar{\sigma}^0 \right)_{\dot{\delta} \alpha} \tilde{S}_{\kappa \epsilon}{\left( x' \right)} \tilde{S}_{\gamma \omega}{\left( x \right)} \left\{ \chi^\epsilon{\left( x' \right)} , \dot{\psi}^\omega{\left( x \right)} \right\} + \notag \displaybreak[3] \\
	&\qquad + 4 \left( \sigma_0 \right)^{\gamma \dot{\delta}} \left( \bar{\sigma}_0 \right)^{\dot{\kappa} \tau} \bar{\dslash{\partial}}_{\dot{\delta} \alpha} \psi^\omega{\left( x \right)} \dslash{\partial}'_{\tau \dot{\beta}} \chi^\epsilon{\left( x' \right)} \left[ \tilde{S}_{\gamma \omega}{\left( x \right)} , \tilde{R}_{\dot{\kappa} \epsilon}{\left( x' \right)} \right] - \notag \displaybreak[3] \\
	&\qquad - 4 \left( \sigma_0 \right)^{\gamma \dot{\delta}} \left( \bar{\sigma}_0 \right)^{\dot{\kappa} \tau} \left( \bar{\sigma}^0 \right)_{\dot{\delta} \alpha} \boldsymbol{\sigma}_{\tau \dot{\beta}} \cdot \boldsymbol{\nabla}' \tilde{R}_{\dot{\kappa} \epsilon}{\left( x' \right)} \tilde{S}_{\gamma \omega}{\left( x \right)} \left\{ \chi^\epsilon{\left( x' \right)} , \dot{\psi}^\omega{\left( x \right)} \right\} - \notag \\
	&\qquad - 4 \left( \sigma_0 \right)^{\gamma \dot{\delta}} \left( \bar{\sigma}_0 \right)^{\dot{\kappa} \tau} \left( \sigma^0 \right)_{\tau \dot{\beta}} \tilde{R}_{\dot{\kappa} \epsilon}{\left( x' \right)} \boldsymbol{\bar{\sigma}}_{\dot{\delta} \alpha} \cdot \boldsymbol{\nabla} \tilde{S}_{\gamma \omega}{\left( x \right)} \left\{ \psi^\omega{\left( x \right)} , \dot{\chi}^\epsilon{\left( x' \right)} \right\} + \notag \\
	&\qquad + m^2 \left( \sigma_0 \right)_\alpha{}^{\dot{\gamma}} \left( \bar{\sigma}_0 \right)_{\dot{\beta}}{}^\kappa \bar{\chi}^{\dot{\omega}}{\left( x \right)} \bar{\psi}^{\dot{\epsilon}}{\left( x' \right)} \left[ \tilde{\bar{S}}_{\dot{\gamma} \dot{\omega}}{\left( x \right)} , \tilde{\bar{R}}_{\kappa \dot{\epsilon}}{\left( x' \right)}  \right] - \notag \\
	&\qquad - 2 i m \left( \sigma_0 \right)_\alpha{}^{\dot{\gamma}} \left( \bar{\sigma}_0 \right)^{\dot{\kappa} \tau} \left( \sigma^0 \right)_{\tau \dot{\beta}} \tilde{\bar{S}}_{\dot{\kappa} \dot{\epsilon}}{\left( x' \right)} \tilde{\bar{S}}_{\dot{\gamma} \dot{\omega}}{\left( x \right)} \left\{ \bar{\chi}^{\dot{\omega}}{\left( x \right)} , \dot{\bar{\psi}}^{\dot{\epsilon}}{\left( x' \right)} \right\} + \notag \\
	&\qquad + 2 i m \left( \sigma_0 \right)^{\gamma \dot{\delta}} \left( \bar{\sigma}_0 \right)_{\dot{\beta}}{}^\kappa \left( \bar{\sigma}^0 \right)_{\dot{\delta} \alpha} \tilde{\bar{R}}_{\kappa \dot{\epsilon}}{\left( x' \right)} \tilde{\bar{R}}_{\gamma \dot{\omega}}{\left( x \right)} \left\{ \bar{\psi}^{\dot{\epsilon}}{\left( x' \right)} , \dot{\bar{\chi}}^{\dot{\omega}}{\left( x \right)} \right\} + \notag \\
	&\qquad + 4 \left( \sigma_0 \right)^{\gamma \dot{\delta}} \left( \bar{\sigma}_0 \right)^{\dot{\kappa} \tau} \bar{\dslash{\partial}}_{\dot{\delta} \alpha} \bar{\chi}^{\dot{\omega}}{\left( x \right)} \dslash{\partial}'_{\tau \dot{\beta}} \bar{\psi}^{\dot{\epsilon}}{\left( x' \right)} \left[ \tilde{\bar{R}}_{\gamma \dot{\omega}}{\left( x \right)} , \tilde{\bar{S}}_{\dot{\kappa} \dot{\epsilon}}{\left( x' \right)} \right] - \notag \\
	&\qquad - 4 \left( \sigma_0 \right)^{\gamma \dot{\delta}} \left( \bar{\sigma}_0 \right)^{\dot{\kappa} \tau} \left( \bar{\sigma}^0 \right)_{\dot{\delta} \alpha} \boldsymbol{\sigma}_{\tau \dot{\beta}} \cdot \boldsymbol{\nabla}' \tilde{\bar{S}}_{\dot{\kappa} \dot{\epsilon}}{\left( x' \right)} \tilde{\bar{R}}_{\gamma \dot{\omega}}{\left( x \right)} \left\{ \bar{\psi}^{\dot{\epsilon}}{\left( x' \right)} , \dot{\bar{\chi}}^{\dot{\omega}}{\left( x \right)} \right\} - \notag \\
	&\qquad \left. - 4 \left( \sigma_0 \right)^{\gamma \dot{\delta}} \left( \bar{\sigma}_0 \right)^{\dot{\kappa} \tau} \left( \sigma^0 \right)_{\tau \dot{\beta}} \tilde{\bar{S}}_{\dot{\kappa} \dot{\epsilon}}{\left( x' \right)} \boldsymbol{\bar{\sigma}}_{\dot{\delta} \alpha} \cdot \boldsymbol{\nabla} \tilde{\bar{R}}_{\gamma \dot{\omega}}{\left( x \right)} \left\{ \bar{\chi}^{\dot{\omega}}{\left( x \right)} , \dot{\bar{\psi}}^{\dot{\epsilon}}{\left( x' \right)} \right\} \right) \, .
\end{align}
Inserting the previously derived results for the commutation and anticommutation relations between the component fields in position space then yields
\begin{align}
2 \dslash{P}_{\alpha \dot{\beta}}
	&= \int \mathrm{d} \mathbf{x} \left(
	- m^2 \left( \sigma_0 \right)_{\alpha \dot{\beta}} \psi_\epsilon{\left( x \right)} \chi^\epsilon{\left( x \right)}
	- m \left( \sigma_0 \right)_\alpha{}^{\dot{\gamma}} \tilde{R}_{\dot{\beta} \epsilon}{\left( x \right)} \tilde{R}_{\dot{\gamma}}{}^\epsilon{\left( x \right)} + \right. \notag \\
	&\qquad + m \left( \bar{\sigma}_0 \right)_{\dot{\beta}}{}^\kappa \tilde{S}_\kappa{}^\omega{\left( x \right)} \tilde{S}_{\alpha \omega}{\left( x \right)}
	- 4 \left( \sigma_0 \right)^{\gamma \dot{\delta}} \bar{\dslash{\partial}}_{\dot{\delta} \alpha} \psi_\epsilon{\left( x \right)} \dslash{\partial}_{\gamma \dot{\beta}} \chi^\epsilon{\left( x \right)} + \notag \\
	&\qquad + 2 i \left( \bar{\sigma}_0 \right)^{\dot{\kappa} \tau} \boldsymbol{\sigma}_{\tau \dot{\beta}} \cdot \boldsymbol{\nabla} \tilde{R}_{\dot{\kappa}}{}^\omega{\left( x \right)} \tilde{S}_{\alpha \omega}{\left( x \right)}
	+ 2 i \left( \sigma_0 \right)^{\gamma \dot{\delta}} \tilde{R}_{\dot{\beta} \epsilon}{\left( x \right)} \boldsymbol{\bar{\sigma}}_{\dot{\delta} \alpha} \cdot \boldsymbol{\nabla} \tilde{S}_\gamma{}^\epsilon{\left( x \right)} + \notag \\
	&\qquad + m^2 \left( \sigma_0 \right)_{\alpha \dot{\beta}} \bar{\chi}_{\dot{\epsilon}}{\left( x \right)} \bar{\psi}^{\dot{\epsilon}}{\left( x \right)}
	+ m \left( \sigma_0 \right)_\alpha{}^{\dot{\gamma}} \tilde{\bar{S}}_{\dot{\beta} \dot{\epsilon}}{\left( x \right)} \tilde{\bar{S}}_{\dot{\gamma}}{}^{\dot{\epsilon}}{\left( x \right)} - \notag \\
	&\qquad - m \left( \bar{\sigma}_0 \right)_{\dot{\beta}}{}^\kappa \tilde{\bar{R}}_\kappa{}^{\dot{\omega}}{\left( x \right)} \tilde{\bar{R}}_{\alpha \dot{\omega}}{\left( x \right)}
	+ 4 \left( \sigma_0 \right)^{\gamma \dot{\delta}} \bar{\dslash{\partial}}_{\dot{\delta} \alpha} \bar{\chi}_{\dot{\epsilon}}{\left( x \right)} \dslash{\partial}_{\gamma \dot{\beta}} \bar{\psi}^{\dot{\epsilon}}{\left( x \right)} - \notag \\
	&\qquad \left. - 2 i \left( \bar{\sigma}_0 \right)^{\dot{\kappa} \tau} \boldsymbol{\sigma}_{\tau \dot{\beta}} \cdot \boldsymbol{\nabla} \tilde{\bar{S}}_{\dot{\kappa}}{}^{\dot{\omega}}{\left( x \right)} \tilde{\bar{R}}_{\alpha \dot{\omega}}{\left( x \right)}
	- 2 i \left( \sigma_0 \right)^{\gamma \dot{\delta}} \tilde{\bar{S}}_{\dot{\beta} \dot{\epsilon}}{\left( x \right)} \boldsymbol{\bar{\sigma}}_{\dot{\delta} \alpha} \cdot \boldsymbol{\nabla} \tilde{\bar{R}}_\gamma{}^{\dot{\epsilon}}{\left( x \right)} \right) \, .
\end{align}
To extract the Hamiltonian from the momentum operator it has to be contracted with the appropriate Pauli matrix
\begin{align}
\mathcal{H}
	&= \frac{1}{2} \left( \sigma_0 \right)^{\alpha \dot{\beta}} \dslash{P}_{\alpha \dot{\beta}} \, .
\end{align}
The Hamiltonian is therefore given by
\begin{align}
\mathcal{H}
	&= \frac{1}{4} \int \mathrm{d} \mathbf{x} \left(
	2 m^2 \psi{\left( x \right)} \chi{\left( x \right)}
	+ m \tilde{R}_{\dot{\beta} \epsilon}{\left( x \right)} \tilde{R}^{\dot{\beta} \epsilon}{\left( x \right)}
	- m \tilde{S}^{\alpha \omega}{\left( x \right)} \tilde{S}_{\alpha \omega}{\left( x \right)} - \right. \notag \\
	&\qquad - 4 \left( \sigma_0 \bar{\sigma}^\mu \sigma_0 \right)^{\gamma \dot{\beta}} \partial_\mu \psi_\epsilon{\left( x \right)} \dslash{\partial}_{\gamma \dot{\beta}} \chi^\epsilon{\left( x \right)} 
	+ 2 i \left( \bar{\sigma}_0 \sigma^i \bar{\sigma}_0 \right)^{\dot{\kappa} \alpha} \partial_i \tilde{R}_{\dot{\kappa}}{}^\omega{\left( x \right)} \tilde{S}_{\alpha \omega}{\left( x \right)} + \notag \\
	&\qquad + 2 i \left( \sigma_0 \bar{\sigma}^i \sigma_0 \right)^{\gamma \dot{\beta}} \tilde{R}_{\dot{\beta} \epsilon}{\left( x \right)} \partial_i \tilde{S}_\gamma{}^\epsilon{\left( x \right)}
	+ 2 m^2 \bar{\chi}{\left( x \right)} \bar{\psi}{\left( x \right)}
	- m \tilde{\bar{S}}_{\dot{\beta} \dot{\epsilon}}{\left( x \right)} \tilde{\bar{S}}_{\dot{\gamma}}{}^{\dot{\beta} \dot{\epsilon}}{\left( x \right)} + \notag \\
	&\qquad + m \tilde{\bar{R}}^{\alpha \dot{\omega}}{\left( x \right)} \tilde{\bar{R}}_{\alpha \dot{\omega}}{\left( x \right)}
	+ 4 \left( \sigma_0 \bar{\sigma}^\mu \sigma_0 \right)^{\gamma \dot{\beta}} \partial_\mu \bar{\chi}_{\dot{\epsilon}}{\left( x \right)} \dslash{\partial}_{\gamma \dot{\beta}} \bar{\psi}^{\dot{\epsilon}}{\left( x \right)} - \notag \\
	&\qquad \left. - 2 i \left( \bar{\sigma}_0 \sigma^i \bar{\sigma}_0 \right)^{\dot{\kappa} \alpha} \partial_i \tilde{\bar{S}}_{\dot{\kappa}}{}^{\dot{\omega}}{\left( x \right)} \tilde{\bar{R}}_{\alpha \dot{\omega}}{\left( x \right)}
	- 2 i \left( \sigma_0 \bar{\sigma}^i \sigma_0 \right)^{\gamma \dot{\beta}} \tilde{\bar{S}}_{\dot{\beta} \dot{\epsilon}}{\left( x \right)}  \partial_i \tilde{\bar{R}}_\gamma{}^{\dot{\epsilon}}{\left( x \right)} \right) \, .
\end{align}
This expression for the Hamiltonian can be further simplified using relations (\ref{threesigplusthreesig}) and (\ref{threebarsigplusthreebarsig}) in Appendix \ref{Asigmarelations} for the special case where the first and last index are 0
\begin{align}
\sigma^0 \bar{\sigma}^\mu \sigma^0
	&= 2 \eta^{\mu 0} \sigma^0 - \sigma^\mu \, , \\
\bar{\sigma}^0 \sigma^\mu \bar{\sigma}^0
	&= 2 \eta^{\mu 0} \bar{\sigma}^0 - \bar{\sigma}^\mu \, .
\end{align}
The Hamiltonian is then reduced to 
\begin{align}
\mathcal{H}
	&= \int \mathrm{d} \mathbf{x} \left(
	2 \dot{\psi}{\left( x \right)} \dot{\chi}{\left( x \right)}
	+ 2 \boldsymbol{\nabla} \psi{\left( x \right)} \cdot \boldsymbol{\nabla} \chi{\left( x \right)}
	+ \frac{m^2}{2} \psi{\left( x \right)} \chi{\left( x \right)}
	+ 2 \dot{\bar{\chi}}{\left( x \right)} \dot{\bar{\psi}}{\left( x \right)} + \right. \notag \\
	&\qquad + 2 \boldsymbol{\nabla} \bar{\chi}{\left( x \right)} \cdot \boldsymbol{\nabla} \bar{\psi}{\left( x \right)}
	+ \frac{m^2}{2} \bar{\chi}{\left( x \right)} \bar{\psi}{\left( x \right)}
	+ \frac{m}{4} \mathrm{Tr}{\left( \tilde{R}^T{\left( x \right)} \tilde{R}{\left( x \right)} \right)}
	+ \frac{m}{4} \mathrm{Tr}{\left( \tilde{S}^T{\left( x \right)} \tilde{S}{\left( x \right)} \right)} - \notag \\
	&\qquad - i \mathrm{Tr}{\left( \tilde{R}^T{\left( x \right)} \boldsymbol{\bar{\sigma}} \cdot \boldsymbol{\nabla} \tilde{S}{\left( x \right)} \right)}
	+ \frac{m}{4} \mathrm{Tr}{\left( \tilde{\bar{R}}^T{\left( x \right)} \tilde{\bar{R}}{\left( x \right)} \right)}
	+ \frac{m}{4} \mathrm{Tr}{\left( \tilde{\bar{S}}^T{\left( x \right)} \tilde{\bar{S}}{\left( x \right)} \right)} - \notag \\
	&\qquad \left. - i \mathrm{Tr}{\left( \tilde{\bar{S}}^T{\left( x \right)} \boldsymbol{\bar{\sigma}} \cdot \boldsymbol{\nabla} \tilde{\bar{R}}{\left( x \right)} \right)} \right) \, .
\label{Hxspace}
\end{align}
It contains the sum of unbarred spinor products and their barred counterparts which is only restricted to be real but could, at least in principle, be either positive or negative. Therefore, on the first glance it seems that this Hamiltonian could have negative eigenvalues. However, as the Lagrangian is by construction supersymmetric and in addition the Hamiltonian was derived using the supersymmetry algebra the eigenvalues of the Hamiltonian must be positive definite. This can also be shown by deriving the momentum space expansion of the component fields in position space, calculating the commutation and anticommutation relations of the momentum space operators, and determining the normal ordered Hamiltonian in momentum space.

\subsection{The Hamiltonian from Canonical Quantisation}
\label{CHcanonicalquant}
The derivation of the Hamiltonian using the supersymmetry algebra is by construction positive definite and is founded in the fundamental properties of the algebra. However, it immediately raises the question whether this approach is equivalent to a construction of the Hamiltonian from canonical quantisation which doesn't require the Lagrangian to be supersymmetric.

For brevity the discussion is restricted to the Lagrangian without hermitian conjugate contribution. The Hamiltonian from canonical quantisation is then defined as 
\begin{align}
\mathcal{H}_{c.q.}
	&= \int \mathrm{d}^3 \mathbf{x} \left( - \frac{\partial \mathcal{L}}{\partial \dot{\chi}^\tau} \dot{\chi}^\tau
	- \frac{\partial \mathcal{L}}{\partial \dot{\psi}^\tau} \dot{\psi}^\tau
	+ \frac{\partial \mathcal{L}}{\partial \dot{\tilde{S}}^{\tau \omega}} \dot{\tilde{S}}^{\tau \omega}
	+ \frac{\partial \mathcal{L}}{\partial \dot{\tilde{R}}^{\dot{\tau} \omega}} \dot{\tilde{R}}^{\dot{\tau} \omega}
	- \mathcal{L} \right) \, .
\end{align}
Inserting the Lagrangian into this definition of the Hamiltonian results in
\begin{align}
\mathcal{H}_{c.q.}
	&= \int \mathrm{d}^3 \mathbf{x} \left( 2 \dot{\chi}{\left( x \right)} \dot{\psi}{\left( x \right)}
	+ 2 \boldsymbol{\nabla} \chi{\left( x \right)} \boldsymbol{\nabla} \psi{\left( x \right)}
	+ \frac{m^2}{2} \psi{\left( x \right)} \chi{\left( x \right)} 
	- i \mathrm{Tr}{\left( \tilde{R}^T{\left( x \right)} \boldsymbol{\bar{\sigma}} \cdot \boldsymbol{\nabla} \tilde{S}{\left( x \right)} \right)} \right. \notag \\
	&\qquad \left. + \frac{m}{4} \mathrm{Tr}{\left( \tilde{S}^T{\left( x \right)} \tilde{S}{\left( x \right)} \right)}
	+ \frac{m}{4} \mathrm{Tr}{\left( \tilde{R}^T{\left( x \right)} \tilde{R}{\left( x \right)} \right)} \right) \, .
\end{align}
It turns out that the Hamiltonian derived from canonical quantisation after normal ordering is identical to the one derived using the supersymmetry algebra. This is intriguing as it paves the way for a significantly simplified derivation of the Hamiltonian in position space involving fermionic fields with mass dimension one. It represents an extension of the commonly used formalism of canonical quantisation to component fields with non-standard mass dimensions.

\section{Summary}
\label{Csummary}
The primary objective of this article was to construct a supersymmetric model for fermionic fields with mass dimension one.

To achieve this goal it was investigated whether it is possible to obtain a model based on the general scalar superfield commonly used in supersymmetric models. It has been shown that such a model cannot be formulated due to problems constructing a Lagrangian containing kinetic terms for the fermionic fields with mass dimension one. This eliminated all but the trivial solution which corresponds to a constant non-dynamic background spinor field and is not appealing. In addition no consistent second quantisation for the component fields can be constructed.

This motivated the formulation of a model for fermionic fields with mass dimension one based on a general spinor superfield. Up to now no explicit calculations for the general spinor superfield exist in the literature, therefore, necessitating the derivation of the model from the ground up. This included the calculation of all chiral and anti-chiral superfields up to third order in covariant derivatives. To second oder in covariant derivatives there is one chiral and one anti-chiral spinor field while to third order there is one chiral and one anti-chiral second rank spinor field. Interestingly, the chiral second-rank spinor field admits a special case that leads to a scalar superfield while the anti-chiral second-rank spinor field can at most be written as a vector superfield.

Dimensional analysis revealed that there is a large number of possible contributions to the mass and kintic terms. Therefore, the discussion was restricted to terms built from chiral and anti-chiral superfields. The resulting on-shell Lagrangian depends solely on two spinor fields and two second-rank spinor fields which corresponds to 8 fermionic and 8 bosonic degrees of freedom.

As it was not ad hoc clear that the Hamiltonian can be derived from the Lagrangian by canonical quantisation a conservative approach based on the supersymmetry algebra was utilised. It provides an anticommutation relation between the supersymmetry generators which is proportional to the momentum operator that contains the Hamiltonian as 0-th component. This is then related to the Lagrangin via the position space representation of the generators that are proportional to the spacetime integral of the supercurrent which itself can be derived from the Lagrangian. This process ensures a Hamiltonian that is consistent with the initial on-shell Lagrangian as well as the supersymmetry algebra. Therefore, the resulting Hamiltonian is positive definite. 

Subsequently it was shown that the Hamiltonian derived by canonical quantisation is identical to the one calculated using the supersymmetry algebra. This shows that it is possible to extend the commonly used formalism of canonical quantisation to component fields with non-standard mass dimensions.

\appendix
\section{Mathematical Appendix}
\label{ASMath}
The following mathematical Appendix is separated into three subsections. The first subsection summarises general definitions. This is followed by a subsection outlining relations between Grassmann variables while the last subsection provides a collection of relations between $\sigma^\mu$- and $\sigma^{\mu \nu}$-matrices.

\subsection{Conventions}
\label{ASSConventions}
The metric was chosen to be
\begin{eqnarray}
\eta_{\mu \nu}
	&=& \mathrm{diag}{\left( + , - , - , - \right)} \, .
\label{metric}
\end{eqnarray}
while the four dimensional anisymmetic tensor was chosen such that
\begin{eqnarray}
\epsilon^{0 1 2 3}
	&=& - \epsilon_{0 1 2 3}
	= 1 \, .
\end{eqnarray}
The components of two dimensional antisymmetric tensors with dotted and undotted indices are defined as follows
\begin{eqnarray}
\epsilon_{1 2}
	&=& \epsilon^{1 2}
	= - \epsilon_{\dot{1} \dot{2}}
	= - \epsilon^{\dot{1} \dot{2}}
	= 1 \, .
\end{eqnarray}
These tensors can then be used to raise and lower the indices of spinors and tensors in the following way
\begin{eqnarray}
\psi^\alpha
	&=& \epsilon^{\alpha \beta} \psi_\beta \, ,\\
\bar{\psi}^{\dot{\alpha}}
	&=& \bar{\psi}_{\dot{\beta}} \epsilon^{\dot{\beta} \dot{\alpha}} \, .
\end{eqnarray}
The two dimensional epsilon tensors with mixed index structure, one lower and one upper index are then proportional to the Kronecker-$\delta$
\begin{eqnarray}
\epsilon_\alpha{}^\beta
	&=& - \epsilon^\beta{}_\alpha
	= \delta^\beta_\alpha \, , \\
\epsilon^{\dot{\alpha}}{}_{\dot{\beta}}
	&=& - \epsilon_{\dot{\beta}}{}^{\dot{\alpha}}
	= \delta^{\dot{\alpha}}_{\dot{\beta}} \, .
\end{eqnarray}
Finally, the $\sigma$-matrices with two Lorentz indices are defined as
\begin{eqnarray}
\sigma^{\mu \nu}
	&= i \sigma^\mu \bar{\sigma}^\nu - i \eta^{\mu \nu} \, ,\\
\bar{\sigma}^{\mu \nu}
	&= i \bar{\sigma}^\mu \sigma^\nu - i \eta^{\mu \nu} \, ,
\end{eqnarray}
where the Pauli matrices are given by
\begin{eqnarray}
\sigma^0
	&= \begin{pmatrix} 1 & 0 \\ 0 & 1 \end{pmatrix} ,
\sigma^1
	= \begin{pmatrix} 0 & 1 \\ 1 & 0 \end{pmatrix} ,
\sigma^2
	= \begin{pmatrix} 0 & - i \\ i & 0 \end{pmatrix} ,
\sigma^3
	= \begin{pmatrix} 1 & 0 \\ 0 & - 1 \end{pmatrix} .
\end{eqnarray}

\subsection{Relations between Grassmann Variables}
\label{SSGrassmann}
It can be shown that the product of two dotted or undotted Grassmann variables with different spinor indices are always proportional to a two dimensional epsilon-tensor with respectively dotted or undotted indices
\begin{eqnarray}
\theta^\alpha \theta^\beta
	&=& - \frac{1}{2} \epsilon^{\alpha \beta} \theta^2 \, , \\
\bar{\theta}^{\dot{\alpha}} \bar{\theta}^{\dot{\beta}}
	&=& - \frac{1}{2} \epsilon^{\dot{\alpha} \dot{\beta}} \bar{\theta}^2 \, .
\end{eqnarray}
Sometimes it also proves useful to rewrite the product of one unbarred and one barred Grassmann variable
\begin{eqnarray}
\theta^\alpha \bar{\theta}^{\dot{\beta}}
	&=& \frac{1}{2} \theta \sigma^\mu \bar{\theta} \left( \sigma_\mu \right)^{\alpha \dot{\beta}} \, .
\end{eqnarray}

\subsection{Relations between $\sigma$-matrices}
\label{Asigmarelations}
A very good source for relations between $\sigma$-matrices can be found in the appendix of \cite{wess82}. However, it is necessary to determine the appropriate phase factors as their choice of conventions for the metric and Dirac matrices differs from the one used in this thesis.

In the following section numerous relations involving two, three, or four $\sigma$-matrices as well as relations involving $\sigma^{\mu \nu}$ are summarised.
\begin{align}
\left( \bar{\sigma}^\mu \right)^{\dot{\gamma} \alpha} \left( \sigma^\nu \right)_{\alpha \dot{\gamma}}
		&= \left( \bar{\sigma}^\mu \sigma^\nu \right)^{\dot{\gamma}}{}_{\dot{\gamma}}
		= \text{Tr}{\left( \bar{\sigma}^\mu \sigma^\nu \right)}
		= 2 \eta^{\mu \nu} \\
\left( \sigma^\mu \right)_{\alpha \dot{\gamma}} \left( \bar{\sigma}^\nu \right)^{\dot{\gamma} \alpha}
		&= \left( \sigma^\mu \bar{\sigma}^\nu \right)_\alpha{}^\alpha
		= \text{Tr}{\left( \sigma^\mu \bar{\sigma}^\nu \right)}
		= 2 \eta^{\mu \nu}
\end{align}

\begin{align}
\left( \sigma^\mu \right)_{\alpha \dot{\beta}} \left( \bar{\sigma}_\mu \right)^{\dot{\gamma} \delta}
	&= - 2 \epsilon_\alpha{}^\delta \epsilon_{\dot{\beta}}{}^{\dot{\gamma}}
	= 2 \delta_\alpha^\delta \delta_{\dot{\beta}}^{\dot{\gamma}} \\
\left( \sigma^\mu \right)_{\alpha \dot{\beta}} \left( \bar{\sigma}_\mu \right)^{\dot{\beta} \alpha}
	&= 2 \delta_\alpha^\alpha \delta_{\dot{\beta}}^{\dot{\beta}} 
	= 8
\end{align}
\begin{align}
\left( \bar{\sigma}^\mu \sigma^\nu + \bar{\sigma}^\nu \sigma^\mu \right)^{\dot{\alpha}}{}_{\dot{\beta}}
	&= 2 \eta^{\mu \nu} \delta^{\dot{\alpha}}_{\dot{\beta}} \\
\left( \sigma^\mu \bar{\sigma}^\nu + \sigma^\nu \bar{\sigma}^\mu \right)_\alpha{}^\beta
	&= 2 \eta^{\mu \nu} \delta_\alpha^\beta
\end{align}
\begin{align}
\sigma^\mu \bar{\sigma}^\nu \sigma^\rho + \sigma^\rho \bar{\sigma}^\nu \sigma^\mu
	&= 2 \eta^{\nu \rho} \sigma^\mu - 2 \eta^{\mu \rho} \sigma^\nu + 2 \eta^{\mu \nu} \sigma^\rho
\label{threesigplusthreesig} \\
\bar{\sigma}^\mu \sigma^\nu \bar{\sigma}^\rho + \bar{\sigma}^\rho \sigma^\nu \bar{\sigma}^\mu
	&= 2 \eta^{\nu \rho} \bar{\sigma}^\mu - 2 \eta^{\mu \rho} \bar{\sigma}^\nu + 2 \eta^{\mu \nu} \bar{\sigma}^\rho
\label{threebarsigplusthreebarsig}
\end{align}
\begin{align}
\sigma^\mu \bar{\sigma}^\nu \sigma^\rho - \sigma^\rho \bar{\sigma}^\nu \sigma^\mu
	&= - 2 i \epsilon^{\mu \nu \rho \tau} \sigma_\tau \\
\bar{\sigma}^\mu \sigma^\nu \bar{\sigma}^\rho - \bar{\sigma}^\rho \sigma^\nu \bar{\sigma}^\mu
	&= 2 i \epsilon^{\mu \nu \rho \tau} \bar{\sigma}_\tau
\end{align}

\begin{align}
\mathrm{Tr}{\left( \sigma^{\mu \nu}\right)}
	&= 0 \\
\mathrm{Tr}{\left( \bar{\sigma}^{\mu \nu}\right)}
	&= 0
\end{align}
\begin{align}
\mathrm{Tr}{\left( \sigma^\mu \bar{\sigma}^\nu \sigma^\rho \bar{\sigma}^\sigma \right)}
	&= 2 \eta^{\rho \sigma} \eta^{\mu \nu}
	- 2 \eta^{\nu \sigma} \eta^{\mu \rho}
	+ 2 \eta^{\nu \rho} \eta^{\mu \sigma}
	- 2 i \epsilon^{\mu \nu \rho \sigma} \\
\mathrm{Tr}{\left( \bar{\sigma}^\mu \sigma^\nu \bar{\sigma}^\rho \sigma^\sigma \right)}
	&= 2 \eta^{\rho \sigma} \eta^{\mu \nu}
	- 2 \eta^{\nu \sigma} \eta^{\mu \rho}
	+ 2 \eta^{\nu \rho} \eta^{\mu \sigma}
	+ 2 i \epsilon^{\mu \nu \rho \sigma}
\end{align}
\begin{align}
\mathrm{Tr}{\left( \sigma^{\mu \nu} \sigma^{\rho \sigma} \right)}
	&= 2 \eta^{\nu \sigma} \eta^{\mu \rho}
	- 2 \eta^{\nu \rho} \eta^{\mu \sigma}
	+ 2 i \epsilon^{\mu \nu \rho \sigma} \\
\mathrm{Tr}{\left( \bar{\sigma}^{\mu \nu} \bar{\sigma}^{\rho \sigma} \right)}
	&= 2 \eta^{\nu \sigma} \eta^{\mu \rho}
	- 2 \eta^{\nu \rho} \eta^{\mu \sigma}
	- 2 i \epsilon^{\mu \nu \rho \sigma}
\end{align}

\begin{align}
\left( \sigma^\mu \bar{\sigma}^{\nu \rho} \right)_{\alpha \dot{\beta}}
	&= - i \eta^{\mu \rho}  \left( \sigma^\nu \right)_{\alpha \dot{\beta}}
	+ i \eta^{\mu \nu} \left( \sigma^\rho \right)_{\alpha \dot{\beta}}
	+ \epsilon^{\mu \nu \rho \sigma} \left( \sigma_\sigma \right)_{\alpha \dot{\beta}} \\
\left( \sigma^{\mu \nu} \sigma^\rho \right)_{\alpha \dot{\beta}}
	&= i \eta^{\nu \rho} \left( \sigma^\mu \right)_{\alpha \dot{\beta}}
	- i \eta^{\mu \rho} \left( \sigma^\nu \right)_{\alpha \dot{\beta}}
	+ \epsilon^{\mu \nu \rho \sigma} \left( \sigma_\sigma \right)_{\alpha \dot{\beta}} \\
\left( \bar{\sigma}^\mu \sigma^{\nu \rho} \right)_{\dot{\alpha} \beta}
	&= - i \eta^{\mu \rho} \left( \bar{\sigma}^\nu \right)_{\dot{\alpha} \beta}
	+ i \eta^{\mu \nu} \left( \bar{\sigma}^\rho \right)_{\dot{\alpha} \beta}
	- \epsilon^{\mu \nu \rho \sigma} \left( \bar{\sigma}_\sigma \right)_{\dot{\alpha} \beta} \\
\left( \bar{\sigma}^{\mu \nu} \bar{\sigma}^\rho \right)_{\dot{\alpha} \beta}
	&= i \eta^{\nu \rho} \left( \bar{\sigma}^\mu \right)_{\dot{\alpha} \beta}
	- i \eta^{\mu \rho} \left( \bar{\sigma}^\nu \right)_{\dot{\alpha} \beta}
	- \epsilon^{\mu \nu \rho \sigma} \left( \bar{\sigma}_\sigma \right)_{\dot{\alpha} \beta}
\end{align}

\begin{align}
\left( \sigma^{\mu \nu} \sigma^{\rho \sigma} \right)_\alpha{}^\beta
	&= - \eta^{\nu \rho} \eta^{\mu \sigma} \epsilon_\alpha{}^\beta
	+ i \eta^{\nu \rho} \left( \sigma^{\mu \sigma} \right)_\alpha{}^\beta
	+ \eta^{\mu \rho} \eta^{\nu \sigma} \epsilon_\alpha{}^\beta
	- i \eta^{\mu \rho} \left( \sigma^{\nu \sigma} \right)_\alpha{}^\beta + \notag \\
	&\quad + i \epsilon^{\mu \nu \rho \sigma} \epsilon_\alpha{}^\beta
	+ \epsilon^{\mu \nu \rho}{}_\tau \left( \sigma^{\tau \sigma} \right)_\alpha{}^\beta
	- i \eta^{\rho \sigma} \left( \sigma^{\mu \nu} \right)_\alpha{}^\beta \\
\left( \bar{\sigma}^{\mu \nu} \bar{\sigma}^{\rho \sigma} \right)^{\dot{\alpha}}{}_{\dot{\beta}}
	&= - \eta^{\nu \rho} \eta^{\mu \sigma} \epsilon^{\dot{\alpha}}{}_{\dot{\beta}}
	+ i \eta^{\nu \rho} \left( \bar{\sigma}^{\mu \sigma} \right)^{\dot{\alpha}}{}_{\dot{\beta}}
	+ \eta^{\mu \rho} \eta^{\nu \sigma} \epsilon^{\dot{\alpha}}{}_{\dot{\beta}}
	- i \eta^{\mu \rho} \left( \bar{\sigma}^{\nu \sigma} \right)^{\dot{\alpha}}{}_{\dot{\beta}} - \notag \\
	&\quad - i \epsilon^{\mu \nu \rho \sigma} \epsilon^{\dot{\alpha}}{}_{\dot{\beta}} 
	- \epsilon^{\mu \nu \rho}{}_\tau \left( \bar{\sigma}^{\tau \sigma} \right)^{\dot{\alpha}}{}_{\dot{\beta}}
	- i \eta^{\rho \sigma} \left( \bar{\sigma}^{\mu \nu} \right)^{\dot{\alpha}}{}_{\dot{\beta}}
\end{align}

\acknowledgments
This work was supported by NSERC Canada.

\end{document}